\documentstyle[12pt]{article}
\newcommand{\mathbf}[1]{\mbox{\boldmath $ #1 $}}
\newcommand{\ie}{{\em i.e.}}    
\newcommand{\veC}[1]{\stackrel{\rightarrow}{#1}}        
\newcommand{\Vec}[1]{\stackrel{\leftarrow}{#1}}         
\newcommand{\fm}{{\rm fm}}              
\newcommand{\MeV}{{\rm MeV}}            
\newcommand{\GeV}{{\rm GeV}}            
\begin{document}
\begin{titlepage}

\noindent{\sl Nuclear Physics A}\hfill LBL-37623\\[8ex]

\begin{center}
{\large {\bf  Treatment of Pionic Modes at the Nuclear Surface \\
              for Transport Descriptions$^*$}}\\[8ex]
{\sl Johan Helgesson and J{\o}rgen Randrup}\\[1ex]
Nuclear Science Division, Lawrence Berkeley Laboratory\\
University of California, Berkeley, California 94720\\[4ex]

August 21, 1995\\[4ex]     
{\sl Abstract:}
\end{center}
Dispersion relations and amplitudes of collective pionic modes
are derived in a $\pi + N N^{-1} + \Delta N^{-1}$ model
for use in transport descriptions
by means of a local density approximation.
It is discussed how pionic modes can be converted to real particles
when penetrating the nuclear surface
and how earlier treatments can be improved.
When the surface is stationary only free pions emerge.
The time-dependent situation is also addressed,
as is the conversion of non-physical
(\ie\ unperturbed $\Delta N^{-1}$) modes
to real particles when the nuclear density vanishes.
A simplified one-dimensional scenario is used to investigate
the reflection and transmission of pionic modes at the nuclear surface.
It is found that reflection of pionic modes is rather unlikely,
but the process can be incorporated into transport descriptions
by the use of approximate local transmission coefficients.

\vfill
{\small\noindent
$^*$This work was supported by the Swedish Natural Science Research Council
    and by the Director, Office of Energy Research, Office of High Energy
    and Nuclear Physics, Nuclear Physics Division of the U.S. Department of
    Energy under Contract No.\ DE-AC03-76SF00098.\\
\noindent
}
\end{titlepage}


\section{Introduction}

In collisions between two heavy nuclei
at bombarding energies from a few hundred MeV
up to several GeV per nucleon,
hadronic matter at high density and temperature is formed.
In such collisions a large number
of energetic particles are produced
and may be used as probes
of the hot and dense phase of the reaction
\cite{Metag,Cassing,Mosel}.

Microscopic transport models,
such as BUU and QMD \cite{Cassing,Mosel,Wolf,Aichelin}
have been fairly successful
in describing particle production in heavy-ion reactions.
In these transport models,
the nucleons propagate in an effective one-body field
while subject to direct two-body collisions.
Sufficiently energetic nucleon-nucleon collisions may agitate
one or both of the colliding nucleons to a nucleon resonance,
especially $\Delta(1232)$, $N^*(1440)$, and $N^*(1535)$.
Such resonances propagate in their own mean field
and may collide with nucleons or other nucleon resonances as well.
Furthermore, the nucleon resonances may decay by meson emission
and these decay processes constitute the main mechanisms
for the production of energetic mesons \cite{Mosel}.

The transport descriptions normally employ
the vacuum properties of the resonances and mesons,
\ie\ the needed cross sections, decay widths, and dispersion relations
are taken according to their values in vacuum \cite{Wolf}.
However, in infinite nuclear matter,
a system of interacting $\pi$ mesons, nucleons, and $\Delta$ isobars
will couple to form spin-isospin modes.
Some of these modes are non-collective in their character,
dominated by a single baryon-hole excitation,
while other modes are collective
and correspond to meson-like states (quasimesons).
The non-collective spin-isospin modes correspond
to those already included in transport simulations
by promoting a nucleon from below to above the Fermi surface.
The collective spin-isospin modes can effectively
be regarded as particles of mesonic character (quasimesons),
which by means of a local density and temperature approximation
can be incorporated into the transport descriptions.

Some in-medium modifications have already
been employed in calculations of heavy-ion collisions,
both qualitatively \cite{Weise}
and by transport simulations \cite{Bertsch,Giessen,Texas}.
A more elaborate $\pi + N N^{-1} + \Delta N^{-1}$ model was
employed in ref.\ \cite{main} to derive several quantities useful for
implementation of in-medium properties in transport descriptions.
While it is straightforward
to apply the local density approximation
in the interior regions of the nuclear system,
there are conceptual problems of how to proceed
at the nuclear surface where the density approaches zero
and the quasimesons convert
to real physical particles.
The problem is that while in a real system
no hole states exist in vacuum,
a collective ($\Delta N^{-1}$-like) mode
can in a stationary infinite system
exist for an arbitrary small (but finite) density.
This is because the particles in the stationary system
have infinite time to explore the entire system
and form collective modes also at extremely low densities.

In this paper we will therefore discuss
how a proper conversion from quasimesons
to real particles at the surface
can be performed.
Earlier works \cite{Giessen,Texas} have treated this conversion
by various approximations
(see further the discussion in section \ref{sec_Qpions}).
In this work we will present a somewhat different approach,
based on ref.\ \cite{main},
and discuss how the approximations in the earlier works
can be improved upon.

The present paper constitutes a qualitative investigation
of treatment of pionic modes at a nuclear surface
in transport simulations.
To make the presentation simple and transparent
we will therefore restrict ourselves to some special cases
of the more complete investigation in ref.\ \cite{main}.
We will only consider the collective modes
in the  spin-longitudinal channel (pion like),
since this is the dominant channel
at the energies we have in mind in this paper.
Collective modes
in the spin-transverse channel ($\rho$ meson like)
can be treated completely analogously.
Furthermore we will consider the zero-temperature case,
and the $\Delta$ width will not be included
in the calculation of the dispersion relations
(denoted as the reference case in ref.\ \cite{main}).

The justification for considering only $T=0$ in this paper
is that there is not a strong dependence
on the dispersion relations of the collective modes
for moderately large temperatures,
especially not at low densities at the surface.
Also, the temperature is not expected to be very high at the surface.
However we want to emphasize
that there is no principal difficulty
associated with incorporating $T>0$ in the treatment.

The motivation for omitting the $\Delta$ width,
in the calculation of the dispersion relations,
is somewhat more involved.
Including the $\Delta$ width self-consistently
in the calculation of the spin-isospin modes
encompasses decay processes like
\[
   \tilde{\pi}_j    \rightarrow    \Delta N^{-1}
                    \rightarrow    (N+\tilde{\pi}_k) N^{-1}\ .
\]
However, since such processes are already explicitly contained
in the transport simulation by processes like
\[
   \tilde{\pi}_j + N   \rightarrow    \Delta
                       \rightarrow    N+\tilde{\pi}_k \ ,
\]
it does not seem to be correct
to include the entire self-consistent $\Delta$ width
when calculating the collective modes
to be used in the transport description.
Instead it seems more correct
to use the results obtained with
the $\Delta$ width omitted,
both for the energies of the modes
and for the partial $\Delta$ widths
to be used in the decays $\Delta \rightarrow N +\tilde{\pi}_j$.

However one should note that
by omitting the $\Delta$ width in the dispersion relations
one also fails to take into account the fact
that the pionic modes
have a Breit-Wigner like energy distribution,
analogous with the $\Delta$.
The width and center of this distribution
are determined by the the self-consistent $\Delta$ width
and depends on the particular pionic mode and its momentum.
The center of the distribution approximately corresponds
to the energy found when the $\Delta$ width is omitted \cite{main}.

In section \ref{sec_Model} we will give
a brief presentation of the model.
The dispersion relations and amplitudes
of the spin-isospin modes obtained in infinite nuclear matter,
are presented and discussed in section \ref{sec_DispAmp}.
Section \ref{sec_Qpions} is devoted to a discussion
of the pionic modes at a nuclear surface
and the implications for transport descriptions,
while our results are summarized in section \ref{sec_Sum}.
In addition we present in appendix \ref{sec_Refl}
a discussion of reflection and transmission properties
at the nuclear surface,
and in appendix \ref{sec_RPAsolu}
some technical details for the RPA equations.

\section{The model}					\label{sec_Model}
The model presented in this section
is treated and motivated in detail in ref.\ \cite{main}.
For convenience we here present a brief recapitulation
of the essential points.
Furthermore, the presentation in this section
only treats the spin-longitudinal channel
for the special case
when $T=0$ and the $\Delta$ width is omitted
in the calculation of the spin-isospin modes.

\subsection{Spin-isospin modes in an infinite system}	\label{sec_Model-1}
We consider a system of interacting nucleons ($N$), delta isobars
($\Delta$) and pi mesons ($\pi$).
In order to investigate the in-medium properties of the interacting particles,
we employ a cubic box with side length $L$;
the calculated properties are not sensitive to the actual size,
so we need not take the limit $L\to\infty$ explicitly.

The in-medium properties are obtained by using the Green's function technique,
starting from non-interacting hadrons.
The non-interacting Hamiltonian can be written
\begin{equation}
 H_0 = \sum_k e_k \hat{b}^{\dagger}_k \hat{b}^{ }_k +
   \sum_l \hbar \omega_\pi(\mathbf{q_l})
          \hat{\pi}^{\dagger}_l \hat{\pi}^{ }_l\ .
\label{eq_H0}
\end{equation}
Here the index $k=(\mathbf{p}_k; \, s_k, m_{s_k}; \, t_k, m_{t_k})$
represents the baryon momentum, spin, and isospin.
The spin and isospin quantum numbers, $s_k$ and $t_k$,
take the values $1\over2$ and $3\over2$ for $N$ and $\Delta$, respectively.
The energy of baryon $k$ moving in a (spatially constant) potential
is denoted $e_k$.
The baryon creation and annihilation operators,
$\hat{b}^{\dagger}_k$ and $\hat{b}^{ }_k$,
are normalized such that they satisfy the usual anti-commutation relation,
\begin{equation}
  \{ \hat{b}^{\dagger}_k, \: \hat{b}^{\mbox{ }}_{k'} \} = \delta_{k,k'}\ .
\end{equation}
In the pion part of $H_0$,
the index $l$ represents the pion momentum and isospin,
$l = (\mathbf{p}_l,\lambda_l=0,\pm~1)$.
The meson energy is given by
$\hbar \omega_{\pi} =  [m_{\pi}^2 + \mathbf{q}^2]^{1/2}$
and the creation and annihilation operators of the pion are
normalized such that they satisfy the usual commutation relation,
\begin{equation}
  [ \hat{\pi}^{\dagger}_l, \: \hat{\pi}^{ }_{l'} ] = \delta_{l,l'}\ .
\end{equation}

Note that the $\Delta$ isobar described by $H_0$ has no decay width,
$\Gamma_\Delta=0$.
When the interactions are turned on,
the $\Delta$ width will emerge
and it will then automatically include also the free width.

\subsubsection{Basic interactions}
At the $N \pi N$ and $N \pi \Delta$ vertices we will use effective $p$-wave
interactions, $V_{N \pi N}$ and $V_{N \pi \Delta}$, which in the
momentum representation can be written as \cite{main,OTW}
\begin{eqnarray}
&~&   V_{N \pi N} = i c \: \frac{(\hbar c)^{1\over2}}{L^3} \:
                 \left[ \frac{2m_N c^2}{m_N c^2 + \sqrt{s}} \right]^{1\over2}
                 \frac{f^\pi_{NN}}{m_\pi c^2}\ F_{\pi}(q) \;
                 (\mathbf{\sigma} \mathbf{\cdot} \mathbf{q}_{cm}) \:
                 \veC{\tau} \mathbf{\cdot} \Vec{\phi}_\pi(\mathbf{q})
                 \label{eq_Vnpn}
                  \\
&~&   V_{N \pi \Delta} =
         i c \: \frac{(\hbar c)^{1\over2}}{L^3} \:
         \left[ \frac{2m_\Delta c^2}{m_\Delta c^2 + \sqrt{s}}\right]^{1\over2}
         \frac{f^\pi_{N\Delta}}{m_\pi c^2}\ F_{\pi}(q) \;
         (\mathbf{S^+} \mathbf{\cdot} \mathbf{q}_{cm}) \:
         \veC{T}^+ \mathbf{\cdot} \Vec{\phi}_\pi(\mathbf{q})
         + {\rm h.c.} \phantom{123} \label{eq_Vnpd}
\end{eqnarray}
In these expressions, $\sqrt{s}$ is the center-of-mass energy
in the $N \pi$ system and $\mathbf{q}_{cm}$
is the pion momentum in the $N \pi$ center-of-mass system,
which in the non-relativistic limit is given by
\begin{equation}
  \mathbf{q}_{cm} \approx \frac{m_N c^2}{m_N c^2 + \hbar \omega}\ \mathbf{q}
                  - \frac{\hbar \omega}
                         {m_N c^2 + \hbar \omega}\ \mathbf{p}_N\ ,
\label{eq_qcm}
\end{equation}
where $\hbar \omega$ and $\mathbf{q}$ is the pion energy and momentum, and
$\mathbf{p}_N$ is the nucleon momentum in an arbitrary frame.
The Pauli spin
and isospin matrices are denoted $\mathbf{\sigma}$ and $\vec{\tau}$, and
 $\mathbf{S^+}$ and $\vec{T}^+$ are spin and isospin $1\over2$ to $3\over2$
transition operators normalized such that
$<{3\over2},{3\over2}|S^+_{+1}|{1\over2},{1\over2}>=1$.\footnote{For clarity,
        we generally employ bold-face characters to denote quantities with
        vector and tensor properties under ordinary spatial rotations,
        while arrows are employed to indicate the transformation properties
        under rotations in isospace.}
The momentum representation of the pion field is taken as
\begin{equation}
 \phi^\pi_\lambda(\mathbf{q}) =
    \frac{L^{3/2} \hbar c}{\sqrt{2 \hbar \omega_\pi(\mathbf{q})}}
    \left[ \hat{\pi}^{ }_\lambda(\mathbf{q}) +
           (-1)^\lambda  \hat{\pi}_{-\lambda}^\dagger(-\mathbf{q}) \right]\ .
\end{equation}
The interactions contain a monopole form factor,
\begin{equation}
 F_{\pi}(q) = \frac{\Lambda_\pi^2 - (m_\pi c^2)^2}{\Lambda_\pi^2 - (cq)^2}\ ,
\label{eq_Fpi}
\end{equation}
and the coupling constants are determined at
$(cq)^2 = (\hbar \omega)^2 -  (c \, \mathbf{q})^2 = (m_\pi c^2 )^2$
and
$\sqrt{s} = m_N c^2$ or  $\sqrt{s} = m_\Delta c^2$.

In addition we will also include effective short-range interactions at
nucleon-hole vertices, again written in momentum space,
\begin{equation}
  V_{NN^{-1},NN^{-1}} = \: \left( \frac{\hbar c}{L} \right)^3 \:
                        g_{NN}' \left( \frac{f^\pi_{NN}}{m_\pi} \right)^2
                        |F_g(q)|^2\ (\mathbf{\sigma_1 \cdot \sigma_2})
                        (\veC{\tau_1} \cdot \Vec{\tau_2})\ ,
\label{eq_Vg}
\end{equation}
and the corresponding interactions obtained
when one (or two) of the nucleons is replaced by a $\Delta$.
The strength of the short-range interactions is determined by the
correlation parameters $g_{NN}'$, $g_{N\Delta}'$, and $g_{\Delta \Delta}'$.

\subsection{RPA approximation}
We want to calculate a spin-isospin mode Green's function
within the RPA approximation, symbolically
\begin{equation}
  G^{\rm RPA}(\alpha,\beta;\omega) = G_{0}(\alpha,\beta;\omega) +
     \sum_{\gamma,\kappa} G_{0}(\alpha,\gamma;\omega)\
                          {\cal V}(\gamma,\kappa;\omega)\
                          G^{\rm RPA}(\kappa,\beta;\omega)\ .
\label{eq_GRPA}
\end{equation}
The spin-isospin modes,
here represented by the Green's function $G^{\rm RPA}$,
will in this approximation
be obtained as an infinite iteration of
(non-interacting) pion, nucleon-hole, and $\Delta$-hole states,
represented by the diagonal Green's function $G_0$,
coupled with the interactions specified
in eqs.\ (\ref{eq_Vnpn}--\ref{eq_Vg})
which here are summarized by the symbolic interaction ${\cal V}$.

In nuclear collisions
at beam energies up to about one GeV per nucleon,
which is the domain of application that we have in mind,
only relatively few mesons and isobars are produced
and so the associated quantum-statistical effects may be ignored.
Accordingly, we assume $n_\Delta \ll 1$, and $n_\pi \ll 1$.

A set of RPA equations,
equivalent to eq.\ (\ref{eq_GRPA})
were derived in ref.\ \cite{main}.
{}From these equations eigenvectors and
eigenenergies are obtained
for the different spin-isospin modes.
The eigenvectors will yield the amplitudes
of the different components
($\pi$, $NN^{-1}$, $\Delta N^{-1}$)
forming the particular spin-isospin mode
with the given eigenenergy.
These RPA amplitudes contain important information
about the nature of the different spin-isospin modes.
The spin-isospin modes (or excited RPA states), $|\Psi_\nu>$,
are created by an operator
$Q^{\dagger}_\nu$,
\begin{equation}
  Q^{\dagger}_\nu(\mathbf{q},\lambda) =
        \sum_{jk} X^\nu_{jk}(\mathbf{q},\lambda)
        \hat{b}^{\dagger}_j  \hat{b}^{ }_k +
        \sum_{k}  Z^\nu_k(\mathbf{q},\lambda) \, \hat{\pi}_k^\dagger -
        \sum_{k}  W^\nu_k(\mathbf{q},\lambda) \, \hat{\pi}^{ }_k\ .
\label{eq_Qrpal}
\end{equation}
The quantity
$X^\nu_{jk}(\mathbf{q},\lambda)$
is here the amplitude of the baryon-hole
($N N^{-1}$ or $\Delta N^{-1}$)
component of the spin-isospin mode $|\Psi_\nu>$
at momentum $\mathbf{q}$ and isospin $\lambda$,
while
$Z^\nu_k(\mathbf{q},\lambda)$ and $W^\nu_k(\mathbf{q},\lambda)$,
in the same way, are the amplitudes of the pionic component.
The summation over baryon and meson states
in eq.\ (\ref{eq_Qrpal}) is restricted by taking
$X_{jk} \propto \delta_{\mathbf{p}_j, \mathbf{p}_k + \mathbf{q}}$,
$Z_k \propto \delta_{\mathbf{p}_k, \mathbf{q}} \delta_{\lambda_k, \lambda}$,
and
$W_k \propto \delta_{\mathbf{p}_k, -\mathbf{q}} \delta_{\lambda_k, -\lambda}$.

The RPA equations are obtained from the relation
\begin{equation}
  <[\delta Q,[H,Q^{\dagger}]]> = \hbar \omega <[\delta Q,Q^{\dagger}]>\ ,
\label{eq_RPA-gen}
\end{equation}
with
$\delta Q=\hat{b}^{\dagger}_k\hat{b}^{ }_j$, $\hat{\pi}_r$, or
$\hat{\pi}_r^\dagger$,
and where the brackets $<\cdot>$ denote the expectation value
in the interacting ground state.
It can be shown that the set of RPA solutions
constitutes an orthonormal set.
For convenience the solutions
of the RPA equations of ref.\ \cite{main},
are recapitulated in appendix \ref{sec_RPAsolu}.

\subsubsection{The total $\Delta$ width}

The $\Delta$ self energy $\Sigma_\Delta$ is calculated according to the
diagrams in fig.\ \ref{fig_DseGraph}, by  taking into account all the diagrams
corresponding to the $\Delta$ decaying into a spin-isospin mode and a nucleon,
which then again form a $\Delta$.
In the spin-longitudinal channel we obtain (ref.\ \cite{main})
\begin{equation}
 \Gamma^l_\Delta(E_\Delta,\mathbf{p}_\Delta) = \mbox{Im } \frac{2}{3}
     \left( \frac{\hbar c}{L} \right)^3
     \sum_{\mathbf{q}} \,
     \left[ \theta({\cal E}) - n(\mathbf{p}_\Delta-\mathbf{q}) \right]
     \bar{M}(\Delta N,N \Delta)\ .
\label{eq_DSE2}
\end{equation}
where the energy available for the spin-isospin mode is given by
\begin{equation}
   {\cal E} = E_\Delta - e_N(\mathbf{p}_\Delta-\mathbf{q})\ ,
\end{equation}
and
$\bar{M}(34,12)$ can be expressed as
\begin{equation}
     \bar{M}(34,12)
=
     \sum_{\omega_\nu > 0}
     \left\{ \frac{ h(31;\nu) h(24;\nu) }
                  { \hbar \omega - \hbar \omega_\nu + i \eta }
       -     \frac{ h(31;\nu)  h(24;\nu) }
                  { \hbar \omega + \hbar \omega_\nu - i \eta }
       \right\}
+
    \frac{ f^\pi_{31} f^\pi_{24} }{(m_\pi c^2)^2} F_g^2 g'_{34,12} \ .
\label{eq_Mrpa}
\end{equation}
The factor $h(jk,\nu)$ is obtained
from the interactions at the vertex
consisting of baryons $j$ and $k$,
and the spin-isospin mode $\nu$.
The interactions to be used depend
on the non-interacting states
that the mode consists of
and must therefore be multiplied
by the amplitude of the corresponding state,
\begin{equation}
  h(jk,\nu) \vartheta^l(jk) = \frac{V_{j \pi k}}{\sqrt{2 \hbar \omega_\pi}}
                \, [Z(\nu)+W(\nu)]                     +
                \sum_{mn}  V_{jk,mn} \, X_{mn}(\nu) \ ,
\label{eq_hMotiv}
\end{equation}
where $\vartheta^l(jk)$ is a short hand notation
for the spin-isospin matrix elements
in the spin-longitudinal channel,
$V_{j \pi k}$ is defined in (\ref{eq_Vnpn}) and (\ref{eq_Vnpd}),
$V_{jk,mn}$ is  defined in (\ref{eq_Vg}),
and the amplitudes $X^l_{mn}$, $Z^l$, $W^l$
are defined in eq.\ (\ref{eq_Qrpal}).
The explicit expression for $h(jk,\nu)$
is somewhat lengthy
and has therefore also been relegated to appendix \ref{sec_RPAsolu}.

\subsubsection{Specific $\Delta$ channels}

The total $\Delta$ width
gives the transition probability per unit time
for the $\Delta$ resonance to
decay to any of its decay channels.
In a transport description one explicitly allows
the $\Delta$ resonance to decay
into specific final particles.
Consequently, one needs not only the total $\Delta$ width
(which is the sum of all decay channels)
but also the partial widths governing the decay
into specific RPA channels.
These decay channels consist
of a nucleon and one of the spin-isospin modes.
Since we have access to all the amplitudes
of a given spin-isospin mode
on the different unperturbed states,
it is possible to derive an expression
for the partial contribution to
$\Gamma_\Delta$ from the $\Delta$ decay to a specific mode $\nu$.
The right-hand side of fig.\ \ref{fig_DseGraph}
shows a diagrammatic representation of such a process.
The partial $\Delta$ width for a $\Delta$ decay to a nucleon
and a spin-longitudinal mode $\nu$ becomes \cite{main}
\begin{eqnarray}
  \tilde{\Gamma}_\Delta^\nu(E_\Delta,\mathbf{p}_\Delta)
    & = & \int \frac{d^3p_N}{(2\pi)^3} \,
          \frac{d^3q}{(2\pi)^3} \;
          | \frac{V_{\Delta \pi N}}{\sqrt{2 \hbar \omega_\pi}}
            \cdot [Z(\nu)+W(\nu)] +
            \sum_{mn}  V_{\Delta N,mn} \cdot X_{mn}(\nu)|^2
    \nonumber \\ & ~& \times
            \bar{n}_N(\mathbf{p}_N) \,
            (2\pi)^3 \delta( \mathbf{p}_\Delta - \mathbf{p}_N - \mathbf{q} )
            2\pi \delta( E_\Delta - e_N - \hbar \omega ) \nonumber \\
     & = &  \frac{1}{3} \int \frac{d^3q}{(2\pi)^3} \:
            | h(\Delta N,\omega_\nu) |^2 \,
            \bar{n}_N(\mathbf{p}_\Delta - \mathbf{q}) \,
            2\pi \delta( E_\Delta - e_N - \hbar \omega_\nu )
\label{eq_GnuReal}
\end{eqnarray}
where the factor, $\bar{n}_N = 1-n_N$,
takes into account the Pauli blocking of the nucleon.
Note that when this expression is to be used in transport models
the factor $\bar{n}_N$ should be omitted
since the Pauli blocking is treated explicitly
in the transport description.
The expression (\ref{eq_GnuReal}) is identical
to the contribution from one
of the $\nu$ terms in eq.\ (\ref{eq_DSE2}).

\section{ Dispersion relations and amplitudes }
\label{sec_DispAmp}

{}From eq.\ (\ref{eq_Xx}) in appendix \ref{sec_RPAsolu}
we calculate the energies
of the spin-isospin modes that are formed in the interacting system,
\ie\ their dispersion relations.
Fig.\ \ref{fig_DispT0r10} displays the dispersion relations
at normal nuclear density,
$\rho = \rho^0 = 0.153\ \fm^{-3}$.
In fig.\ \ref{fig_DispT0r10} a number of different modes
in the spin-longitudinal ($\pi$-like) channel are apparent.
Some of those are non-collective $NN^{-1}$ modes (solid curves),
which have their energies within the regions
\begin{eqnarray}
    0   &   \leq   &   \hbar \omega\
            \leq\       \frac{q^2}{2 m_N^*} +  \frac{q p_F}{m_N^*}\ ,
                       \qquad q < 2 p_F                     \ , \nonumber \\
    \frac{q^2}{2 m_N^*} - \frac{q p_F}{m_N^*}
        &   \leq   &   \hbar \omega\
            \leq\       \frac{q^2}{2 m_N^*} +  \frac{q p_F}{m_N^*}\ ,
                       \qquad q > 2 p_F\ .
\label{eq_NhCont}
\end{eqnarray}
Since we are presenting our results for a box normalization
with a finite side length $L$,
we obtain a discrete number of non-collective $NN^{-1}$ modes.
The total number of spin-isospin modes
within the region (\ref{eq_NhCont})
depends on $L$ and tends towards a continuum
in the limit $L \rightarrow \infty$.
Similarly, a number of
non-collective $\Delta N^{-1}$ states emerge
in fig.\ \ref{fig_DispT0r10} which,
for a fixed $\mathbf{q}$, have their energies constrained to a band,
\begin{equation}
    m_\Delta - m_N + \frac{q^2}{2 m_\Delta} - \frac{q p_F}{m_\Delta}\
          \leq\     \hbar\omega\
          \leq\     m_\Delta - m_N + \frac{q^2}{2 m_\Delta} +
                       \frac{q p_F}{m_\Delta}\ .
\label{eq_DhCont}
\end{equation}
The non-collective baryon-hole modes
correspond in a transport description
to propagation of uncoupled baryons ($N$ or $\Delta$).
This was discussed and studied in detail
in ref.\ \cite{main} and will therefore
not be further discussed in this paper.

In addition, two collective modes
appear in fig.\ \ref{fig_DispT0r10},
represented by dot-dashed curves.
The lower one
starts at $\hbar \omega = m_\pi c^2$ at $q = 0$
and continues into the $\Delta N^{-1}$ region
around $q \approx 360\ \MeV/c$.
This mode will in the following be referred to as $\tilde{\pi}_1$.
The upper collective mode
starts slightly above
$\hbar\omega \approx m_\Delta c^2 - m_N c^2$ at $q = 0$
and approaches
$\hbar\omega_\pi = [(m_\pi c^2)^2 + (cq)^2]^{1/2}$
at large $q$.
This mode is denoted $\tilde{\pi}_2$.

The incorporation of the two collective spin-isospin modes
into transport equations is more involved.
These modes can be regarded as separate particles of pionic character,
$\tilde{\pi}_1$ and $\tilde{\pi}_2$,
and treated in a manner analogous
to the standard treatment of the pion.
Since the pion is then fully included in the description,
it should no longer be treated explicitly.
The propagation of the two collective pionic modes
is governed by the effective Hamiltonians
\begin{eqnarray}
  \tilde{H}_1(\mathbf{r},\mathbf{q}) & = &
      \hbar \omega_{1}(\mathbf{q};\rho(\mathbf{r}))\ \equiv\
                  \hbar \tilde{\omega}_1\ ,
  \nonumber \\
  \tilde{H}_2(\mathbf{r},\mathbf{q}) & = &
      \hbar \omega_{2}(\mathbf{q};\rho(\mathbf{r}))\ \equiv\
      \hbar \tilde{\omega}_2\ ,
\label{eq_Hqpion}
\end{eqnarray}
where $\hbar \omega_{1}$ and $\hbar \omega_{2}$
are the energy-momentum relations
for the lower and upper collective modes
displayed in fig.\ \ref{fig_DispT0r10} for $\rho=\rho^0$.
Note that the spatial dependence of $\tilde{H}_1(\mathbf{r},\mathbf{q})$
is incorporated by  representing $\rho(\mathbf{r})$ as a local quantity.
To facilitate center-of-mass transformations we parametrize
the dispersion relations of the pionic modes in the form
\begin{equation}
    \hbar \tilde{\omega}(\mathbf{q};\rho)
\approx
    \left\{  [c \, q - c \, q_0(\rho)]^2 + m_0(\rho)^2 c^4
    \right\}^{\frac{1}{2}} + U_0(\rho)
\label{eq_DispParam}
\end{equation}
For convenience in the transport simulations,
we have chosen to use relatively simple expressions
for the parametrization,
rather than to try to optimize the fit.
In this way a quasipion moves like a relativistic particle
with the group velocity determined
by an effective momentum and energy
\begin{equation}
    \frac{d \hbar \tilde{\omega}}{dq}
=
    c \frac{c(q-q_0)}{\hbar \tilde{\omega} -U_0}
=
    c \frac{c q^*}{\hbar \tilde{\omega}^*}\ .
\end{equation}
The density-dependent parameters $q_0$, $m_0$ and $U_0$
are presented in fig.\ \ref{fig_DispParam}.

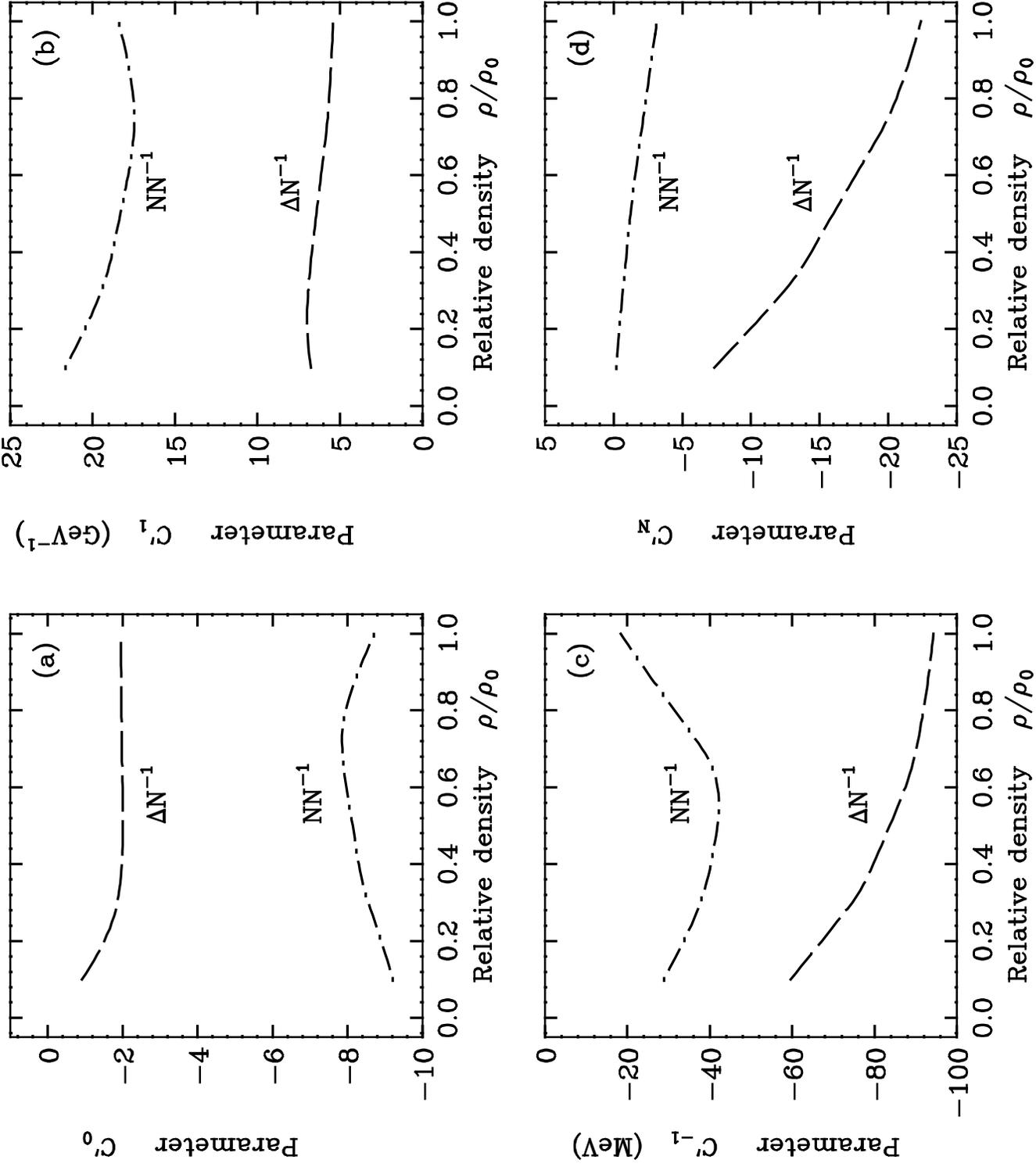
\begin{figure}\caption{}\label{fig_AmpuParam}
The density-dependent parameters,
$C_0'$ (part $a$),
$C_1'$ ($b$),
$C_{-1}'$ ($c$) and
$C_N'$ ($d$),
for the amplitudes
of the $\Delta N^{-1}$ (long-dashed curve)
and the $N N^{-1}$ (dot-dashed) components
on the upper pionic branch, eqs.\ (\ref{eq_XDh-upp-parm})
and (\ref{eq_XNh-upp-parm}).
\end{figure}

Furthermore, in the collision term of the standard transport description
the process for the production and absorption of pions
$\Delta \leftrightarrow N + \pi$,
should be replaced by the two distinct processes
\begin{equation} \Delta
  \leftrightarrow N + \tilde{\pi}_1 \quad \mbox{ and } \quad \Delta
  \leftrightarrow N + \tilde{\pi}_2\ .
\label{eq_Ddecay}
\end{equation}
The $\Delta$ decay is governed
by the $\Delta$ decay width in the medium
to these two specific channels,
$\tilde{\Gamma}_\Delta^j$.
These partial widths,
should be employed in the same manner as the free width,
\ie\ they describe the probability for the $\Delta$ isobar
to decay into a nucleon and a pion.
The only difference is that several collective pionic modes
are available in the final state.
In fig.\ \ref{fig_GamParNPB} we present total and partial
$\Delta$ widths for a $\Delta$ with the momentum 300 MeV/$c$.
The reverse processes in (\ref{eq_Ddecay}) are characterized by
cross sections that were presented and discussed in ref.\ \cite{main}.
To obtain the partial $\Delta$ width from eq.\ (\ref{eq_GnuReal})
it is necessary to know the amplitudes
$Z$, $\sum X_{\Delta N^{-1}}$ and $\sum X_{N N^{-1}}$.
We therefore also present a parametrization of these quantities.
On the lower pionic mode the pionic component
dominates for small momenta, $q$,
and the $\Delta N^{-1}$ component dominates at larger momenta.
Therefore the sum of all individual $\Delta N^{-1}$ components
will for small $q$ increase with $q$.
However, when the lower pionic mode
enters the $\Delta N^{-1}$ region
the collectivity disappears gradually,
and the sum of all individual $\Delta N^{-1}$ components
starts to decrease with $q$.
Thus we employ the form
\begin{eqnarray}
               \sum X^{\tilde{\pi}_1}_{\Delta N^{-1}}(\mathbf{q};\rho)
&  \approx &
               f_\Delta(q,\, \vec{C}[X^{\tilde{\pi}_1}_{\Delta N^{-1}}])\ ,
\label{eq_XDh-low-parm}
\\
               \sum X^{\tilde{\pi}_1}_{N N^{-1}}(\mathbf{q};\rho)
&  \approx  &
               f_N(q,\, \vec{C}[X^{\tilde{\pi}_1}_{ N N^{-1}}])\ ,
\label{eq_XNh-low-parm}
\end{eqnarray}
with
\begin{equation}
    f_{\stackrel{\scriptstyle \Delta}{\scriptstyle N} }(q,\vec{C})
=
               \frac{ \pm C_1^2 + C_2 \, cq }{ C_3^2 + (c q - C_4)^2 }\ .
\label{eq_f1(q,rho)-parm}
\end{equation}
The density-dependent coefficients $\vec{C}$
are presented in fig.\ \ref{fig_AmplParam}.
On the upper pionic mode we instead parametrize the amplitudes as
\begin{eqnarray}
               \sum X^{\tilde{\pi}_2}_{\Delta N^{-1}}(\mathbf{q};\rho)
& \approx &
               C_N'[X^{\tilde{\pi}_2}_{ \Delta N^{-1}}]
               \sqrt{ f_2(q,\, \vec{C}'[X^{\tilde{\pi}_2}_{\Delta N^{-1}}]) }
\label{eq_XDh-upp-parm}
\\
               \sum X^{\tilde{\pi}_2}_{ N N^{-1}}(\mathbf{q};\rho)
& \approx &
               C_N'[X^{\tilde{\pi}_2}_{ N N^{-1}}]
               \sqrt{ 1- f_2(q,\, \vec{C}'[X^{\tilde{\pi}_2}_{ N N^{-1}}]) }\ ,
\label{eq_XNh-upp-parm}
\end{eqnarray}
where
\begin{equation}
    f_2(q,\vec{C}')
=
    \left[ 1 + \exp \left( C_0' + C_1' \, cq  + C_{-1}' / cq \right)
    \right]^{-1}\ ,
\label{eq_f2(q,rho)-parm}
\end{equation}
with the density-dependent coefficients $\vec{C}'$,
displayed in fig.\ \ref{fig_AmpuParam}.
The amplitudes $Z$ of the pion component are obtained from the parametrization
of the squared amplitudes, eqs.\ (\ref{eq_Zl2-parm}) and (\ref{eq_Zu2-parm})
below, as $Z = \sqrt{Z^2}$.

The total $\Delta$ decay width,
has apart from the partial contributions
$\tilde{\Gamma}_\Delta^j$,
also the partial contributions
$\Gamma_\Delta^{N N^{-1}}$ and
$\Gamma_\Delta^{\Delta N^{-1}}$.
The partial width $\Gamma_\Delta^{N N^{-1}}$
gives the probability for the $\Delta$
to decay into a nucleon and a $N N^{-1}$ state.
In a transport description,
this implies that we initially have a $\Delta$
and after the decay process
we have two nucleons above the Fermi surface
and a hole left in the Fermi sea.
But this is the same process
as if the $\Delta$ would collide with a nucleon
below the Fermi surface to give two nucleons above the Fermi surface.
This process is normally already included
in the collision term in a standard transport description,
and the probability for such a collision is given by the
cross section for the process
$\Delta + N \rightarrow N + N$.
In a transport description it is therefore not correct
to both include a $\Delta$ decay according to
$\Gamma_\Delta^{N N^{-1}}$
and a collision term with
$\Delta + N \rightarrow N + N$.
Instead, the correct procedure should be to exclude
$\Gamma_\Delta^{N N^{-1}}$
and modify the cross section
$\sigma(\Delta + N \rightarrow N + N)$
to be the in-medium cross section.
Calculations of such in-medium cross sections was discussed
in ref.\ \cite{main}.
In the same way,
$\Gamma_\Delta^{\Delta N^{-1}}$
should be excluded in a transport description,
and $\sigma(\Delta + N \rightarrow \Delta + N)$
be the in-medium cross section.

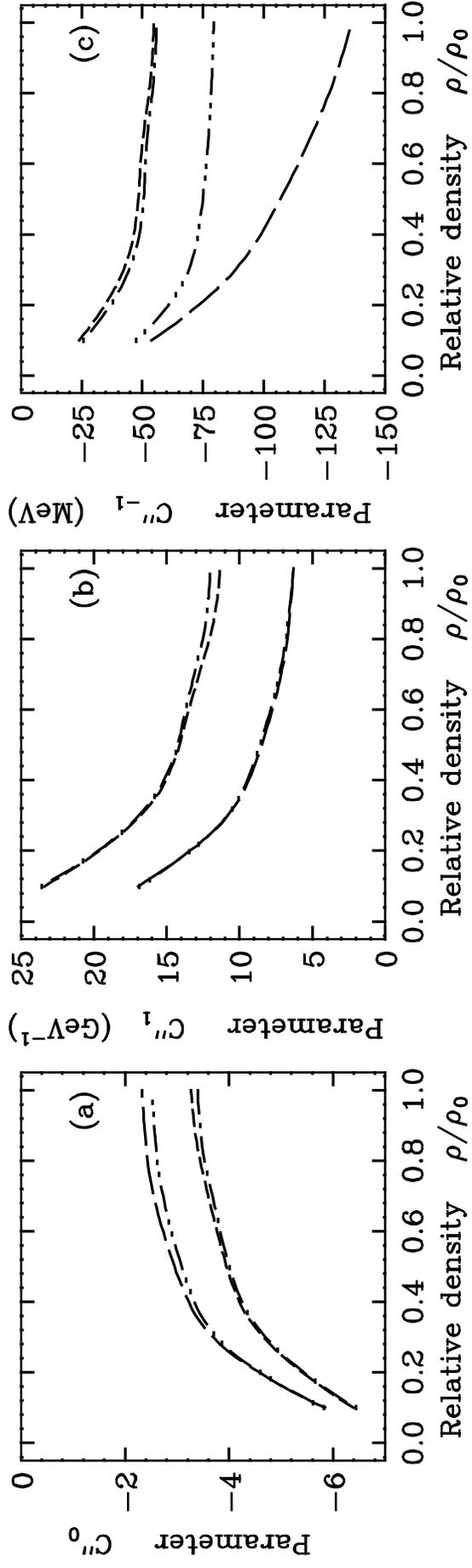
\begin{figure}\caption{}\label{fig_Amp2Param}
The density-dependent parameters,
$C_0''$ (part $a$),
$C_1''$ ($b$), and
$C_{-1}''$ ($c$),
for the squared amplitudes
of the $\pi$ (dot-dashed curve)
and the $\Delta N^{-1}$ (short-dashed) components
on the lower pionic branch,
eqs.\ (\ref{eq_Zl2-parm}) and (\ref{eq_XDhl2-parm})
and of the $\pi$ (dot-dot-dashed)
and the $\Delta N^{-1}$ (long-dashed) components
on the upper pionic branch,
eqs.\ (\ref{eq_Zu2-parm}) and (\ref{eq_XDhu2-parm}).
\end{figure}

Although the collective pionic modes
can thus be effectively treated as ordinary particles,
the fact that their wave functions contain components from
$\pi$, $NN^{-1}$ and $\Delta N^{-1}$ states
makes it difficult to picture them
in a physically simple manner.
Fortunately, their specific structure
is not important for the transport process,
as as long as these quasiparticles remain
well inside the nuclear medium.
First when such a quasiparticle penetrates a nuclear surface
and emerges as a free particle
is it physically meaningful to determine
what kind of real particle it is.
The gradual transformation of the collective quasiparticle
occurs automatically within the formalism,
because as the density is lowered,
$\rho \rightarrow 0$,
a pionic mode will acquire 100\%
of either the pion component or the $\Delta N^{-1}$ component,
depending on $\omega$ and $q$.
That is to say, it will turn into either a free pion
or an unperturbed $\Delta N^{-1}$ state.
The squared amplitudes have their values between zero and unity and we
therefore employ the parametrizations
\begin{eqnarray}
               Z^{\tilde{\pi}_1}(\mathbf{q};\rho)^2
&  \approx  &
               f_2(q,\, \vec{C}''[Z^{\tilde{\pi}_1}]),
\label{eq_Zl2-parm}
\\
               Z^{\tilde{\pi}_2}(\mathbf{q};\rho)^2
& \approx &
               1-f_2(q,\, \vec{C}''[Z^{\tilde{\pi}_2}])
\label{eq_Zu2-parm}
\\
               \sum X^{\tilde{\pi}_1}_{\Delta N^{-1}}(\mathbf{q};\rho)^2
&  \approx  &
               1-f_2(q,\, \vec{C}''[X^{\tilde{\pi}_1}_{\Delta N^{-1}}])
\label{eq_XDhl2-parm}
\\
               \sum X^{\tilde{\pi}_2}_{\Delta N^{-1}}(\mathbf{q};\rho)^2
& \approx &
               f_2(q,\, \vec{C}''[X^{\tilde{\pi}_2}_{\Delta N^{-1}}])\ ,
\label{eq_XDhu2-parm}
\end{eqnarray}
with $f_2(q,\vec{C}'')$ from eq.\ (\ref{eq_f2(q,rho)-parm}),
while the sum of the squared $N N^{-1}$ amplitudes
are obtained from the normalization,
\ie\ all squared amplitudes sum up to unity.
The density-dependent coefficients $\vec{C}''$,
are presented in fig.\ \ref{fig_Amp2Param}.
There remains the practical problem of how to represent
an unperturbed $\Delta N^{-1}$ state when $\rho \rightarrow0$.
However, as will be discussed in section \ref{sec_Qpions},
only a very small fraction  of the pionic modes
(vanishing for a stationary density profile)
will emerge as  unperturbed $\Delta N^{-1}$ states.

Note that this approach is different
from earlier works \cite{Giessen,Texas,KochPriv}
where the nature of the pionic mode
was determined already in the creation process,
\ie\ when the mode was created it was determined whether it represented
a free pion or an unperturbed $\Delta N^{-1}$ state.
This difference in approach will have crucial effects
for the collective modes that escape the system,
as will be seen in the next sections.

\section{ Quasipions at the nuclear surface }
\label{sec_Qpions}

In previous works the quasipions at a surface have been treated
in various approximations.
In ref.\ \cite{Giessen} effective dispersion relations were introduced,
corresponding to modes with either 100\% pionic
or $\Delta N^{-1}$ component.
The pionic mode was then propagated as a quasipion,
emerging as a free pion at the surface,
while the $\Delta N^{-1}$ mode was used
to derive a $\Delta$ potential for the uncoupled $\Delta$s.
In this way only pions and $\Delta$s escape the system,
but the drawback is that the effective dispersion relations
are quite distorted compared to the original ones,
and that the in-medium effects seem to be over-estimated
by allowing all uncoupled $\Delta$s propagate
in a collective potential.

In \cite{Texas} both collective modes were propagated,
and hence some modes escape the system
as unperturbed $\Delta N^{-1}$ states.
This was effectively taken care of
by converting these modes to free $\Delta$s
(neglecting baryon number conservation),
with the justification that the number of modes escaping the system
as unperturbed $\Delta N^{-1}$ states were found to be small.
We will in this paper report on alternative ways
to treat the modes at the surface
based on ref.\ \cite{main},
and how this treatment can improve
the descriptions in refs.\ \cite{Giessen,Texas}.

We first consider the simplified case
when the nuclear surface is stationary
\ie\ $\rho(\mathbf{r},t) = \rho(\mathbf{r},0)$ for all times $t$.
This means that there is no explicit time dependence
in the effective Hamiltonians in equation (\ref{eq_Hqpion}),
and thus the energies of the collective pionic modes are conserved.
For a pionic mode with energy
$\hbar \omega_\nu(\mathbf{q};\rho)$
propagating in a varying density
this means that the momentum
$\mathbf{q}$ will change as $\rho$ changes.
That is to say, the pionic mode effectively feels a potential.
In addition to momentum changes,
also the amplitudes
$X^\nu_{jk}(\mathbf{q};\rho)$,
$Z^\nu(\mathbf{q};\rho)$ and
$W^\nu(\mathbf{q};\rho)$
of the baryon-hole and pion components
will change as the density changes.
In the limit of vanishing density either
$X^\nu_{\Delta N^{-1}}$ or $Z^\nu$ will turn to unity, depending on
the mode $\nu$ and its energy and momentum.

In the latter case no problems arise,
the collective mode has simply been converted
to a free pion escaping the system.
However, in the former case
there is some inconsistency in the formalism
since there are no hole states in vacuum ($\rho=0$).
In a quantal description
of spin-isospin modes propagating at a surface
different scenarios could emerge.
There is some small probability that the mode
could be reflected at the surface.
Alternatively the mode could break up
in an uncoupled $\Delta$ escaping the system,
with the hole is trapped inside the nucleus.
In a transport description this could be handled
by allowing the test particle representing the mode $\nu$
to absorb a nearby nucleon,
converting it into a $\Delta$ isobar.

We consider in this section the quasipions
at a surface propagating  from normal nuclear density to vacuum.
In this case the amplitude $W^\nu$ will be very small
and we will therefore omit it in the qualitative discussion
of this section.

In fig.\ \ref{fig_w12-rho}$a$ we present the energy
$\hbar \tilde{\omega}_1(\mathbf{q})$
of the lower collective mode $\tilde{\pi}_1$
for different densities in the range
$0.1 \rho^0 \leq \rho \leq \rho^0$.
The line closest to the free pion relation (dotted line)
represents the dispersion relation at the lowest density.
As the density is increased
the energy relation is (for each fixed $q$) lowered.
For small $q$ values the mode is completely dominated
by the pion component.
As $q$ is increased also the $\Delta N^{-1}$ components
starts to contribute substantially
(although to different extent depending on the density).
In the approximate range
$300\ \MeV/c \leq q \leq 400\ \MeV/c$
(depending on density)
the mode enters the $\Delta N^{-1}$ continuum
and changes character from collective mode to non-collective.

We will in this section discuss four different examples,
suitably chosen to illustrate the main features.

\subsection{ Lower pionic mode }
\label{sec_QpionsLow}

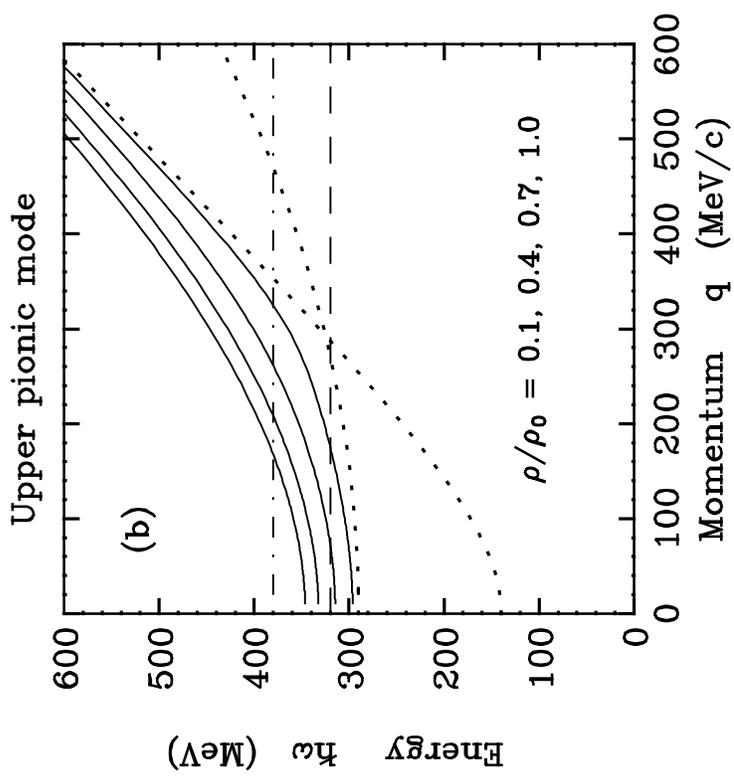
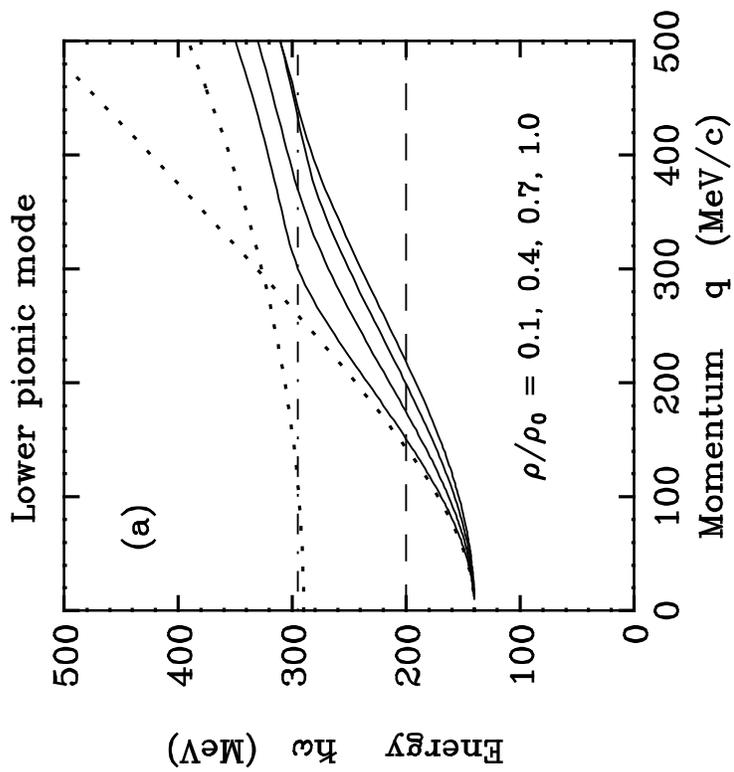
\begin{figure}\caption{}\label{fig_w12-rho}
The energy $\hbar \tilde{\omega}(\mathbf{q})$
for different densities in the range
$0.1 \rho^0 \leq \rho \leq \rho^0$.
In part $a$ is presented the energies
of the lower collective mode $\tilde{\pi}_1$
while the energies of the upper collective mode $\tilde{\pi}_2$
is shown in part $b$.
As references the free pion dispersion relation
and the unperturbed $\Delta N^{-1}$ relation
have been plotted as dotted lines.
The dashed and dot-dashed horizontal lines
indicate the energies considered in the four examples
of sections \ref{sec_QpionsLow} and \ref{sec_QpionsUpp}.
\end{figure}

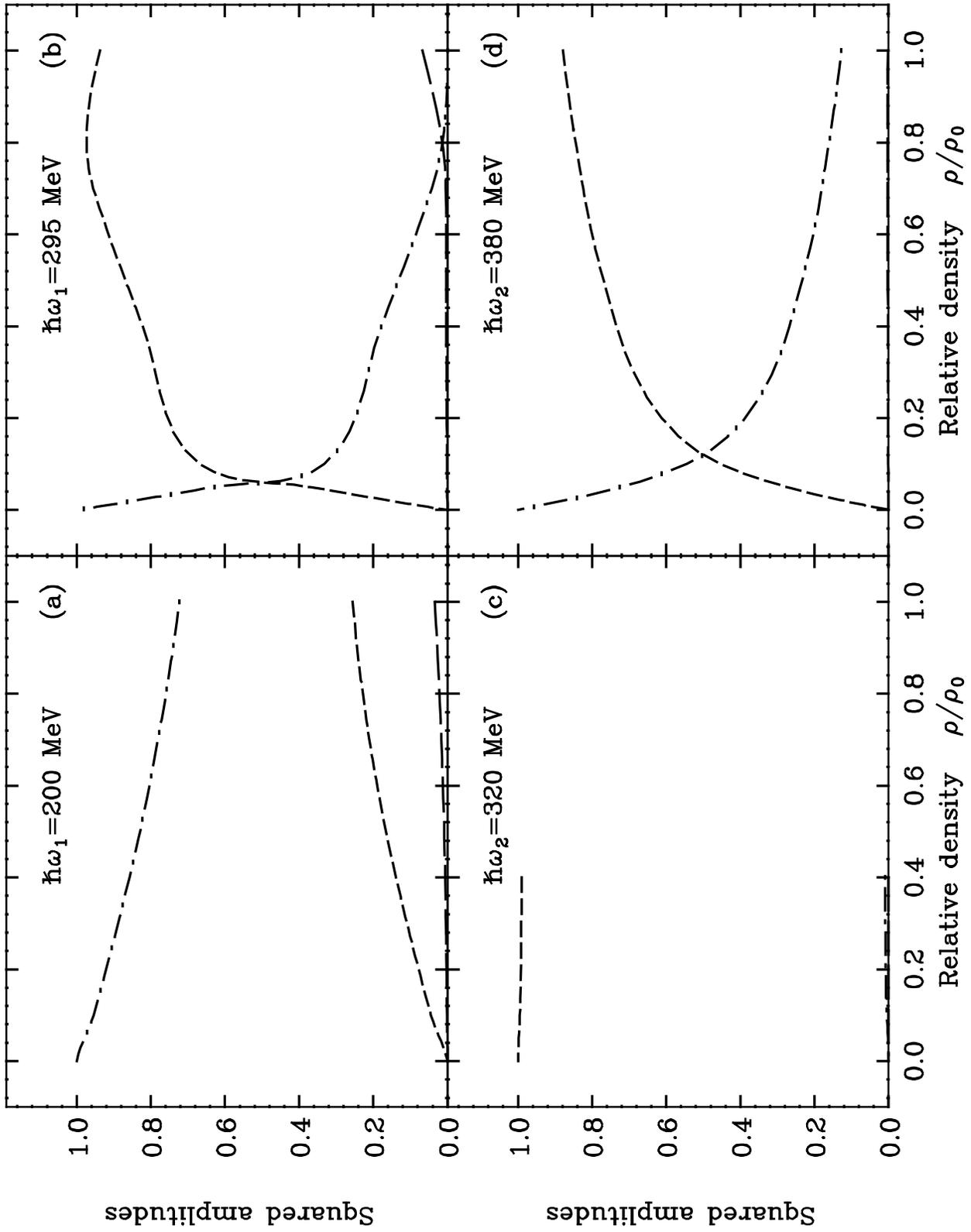
\begin{figure}\caption{}\label{fig_Amp-w}
Sum of all squared amplitudes of the
$\Delta N^{-1}$ and $N N^{-1}$ components,
as well as squared amplitude of the pion component.
In part $a$ and $b$ is shown the squared amplitudes
for the lower pionic mode $\tilde{\pi}_1$
at the energies
$\hbar \tilde{\omega}_1 = 200\ \MeV$
and
$\hbar \tilde{\omega}_1 = 295\ \MeV$,
respectively.
In $c$ and $d$ is displayed the squared amplitudes
for the upper pionic mode $\tilde{\pi}_1$
at the energies
$\hbar \tilde{\omega}_2 = 320\ \MeV$
and
$\hbar \tilde{\omega}_2 = 380\ \MeV$,
respectively.
The dot-dashed curve represents the pion component,
the $\Delta N^{-1}$ component is represented by a short-dashed curve,
and the long-dashed curve represents the $N N^{-1}$ component.
\end{figure}

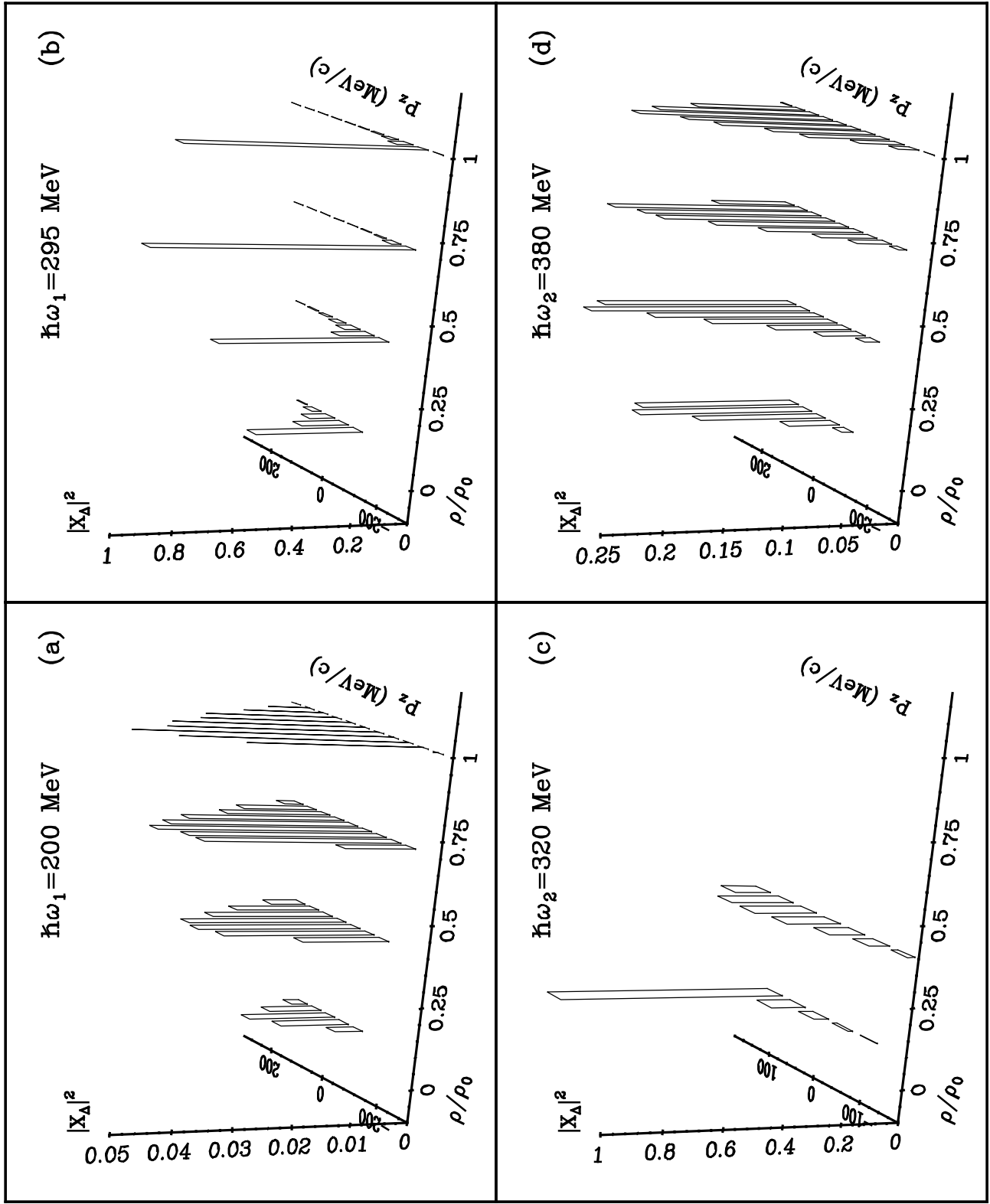
\begin{figure}\caption{}\label{fig_DhAmp-w}
Squared amplitudes of the
individual $\Delta N^{-1}$, components.
In part $a$ and $b$ is shown the squared amplitudes
for the lower pionic mode $\tilde{\pi}_1$
at the energies
$\hbar \tilde{\omega}_1 = 200\ \MeV$
and
$\hbar \tilde{\omega}_1 = 295\ \MeV$,
respectively.
In $c$ and $d$ is displayed the squared amplitudes
for the upper pionic mode $\tilde{\pi}_2$
at the energies
$\hbar \tilde{\omega}_2 = 320\ \MeV$
and
$\hbar \tilde{\omega}_2 = 380\ \MeV$,
respectively.
\end{figure}

In our first example we consider the mode $\tilde{\pi}_1$
created at normal density
with energy $\hbar \tilde{\omega}_1 = 200\ \MeV$
propagating towards vacuum
without any interactions.
Initially this mode has the following characteristics:
\[
\begin{array}{cccc}
q \approx 220\ \MeV/c
&
|Z|^2 \approx 0.72
&
\sum_{\Delta N^{-1}} |X_{\Delta N^{-1}}|^2 \approx 0.25
&
\sum_{N N^{-1}} |X_{N N^{-1}}|^2 \approx 0.03\ .
\end{array}
\]
As the density decreases it will,
due to the energy conservation,
follow the path indicated
by the dashed line in fig.\ \ref{fig_w12-rho}$a$,
from $q \approx 220\ \MeV/c$ to $q \approx 150\ \MeV/c$.
In fig.\ \ref{fig_Amp-w}$a$ we see how the squared amplitudes
vary with the density for this particular energy
of the mode $\tilde{\pi}_1$.
As seen in fig.\ \ref{fig_Amp-w}$a$
only the pion component remains,
as the zero density limit is reached.
Thus in this particular case,
the mode will escape the system as a free pion.
In fig.\ \ref{fig_DhAmp-w}$a$
we show the squared amplitudes
of the individual $\Delta N^{-1}$ amplitudes
for the energy $\hbar \tilde{\omega}_1 = 200\ \MeV$
of the mode $\tilde{\pi}_1$.
We see that the mode is collective
at all densities.

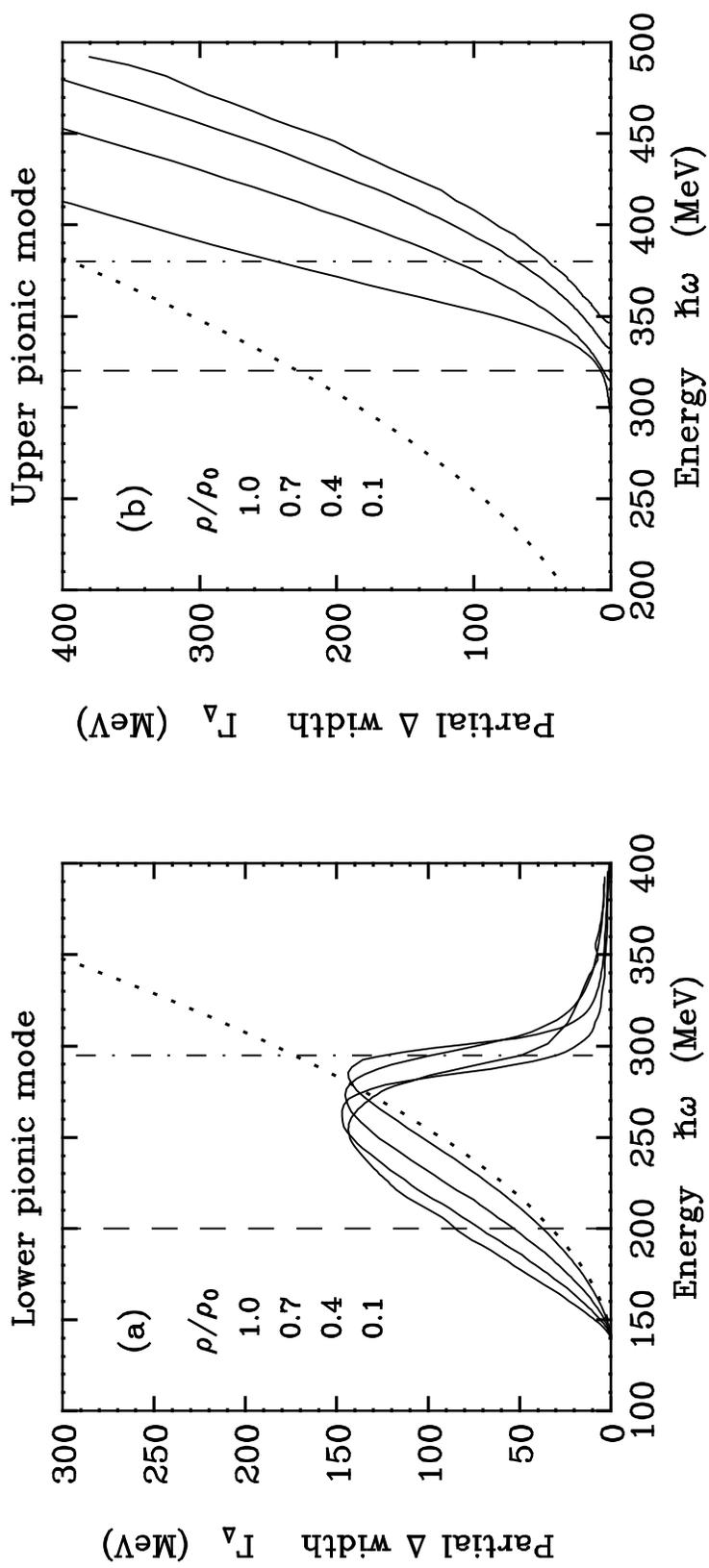
\begin{figure}\caption{}\label{fig_Gam_w-rho}
Partial $\Delta$ decay width for a $\Delta$ at rest,
for different densities in the range
$0.1 \rho^0 \leq \rho \leq \rho^0$.
In part $a$ is presented the partial $\Delta$ width
for the lower collective mode $\tilde{\pi}_1$
while the partial $\Delta$ width
for the upper collective mode $\tilde{\pi}_2$
is shown in part $b$.
As a reference the free $\Delta$ width
has been plotted as a dotted line.
The dashed and dot-dashed vertical lines
indicate the energies considered in the four examples
of sections \ref{sec_QpionsLow} and \ref{sec_QpionsUpp}.
\end{figure}

In the discussion so far
we have not addressed the creation process.
How probable is it that we produce the mode $\tilde{\pi}_1$
at the particular energy $\hbar \tilde{\omega}_1 = 200\ \MeV$?
This information is given
by the partial $\Delta$ decay width
to the mode $\tilde{\pi}_1$.
In fig.\ \ref{fig_Gam_w-rho}$a$ we display
the partial $\Delta$ decay width,
$\tilde{\Gamma}_\Delta^1$,
for different densities,
as a function of the energy
of the emitted quasipion $\tilde{\pi}_1$.
Note that the width displayed
in fig.\ \ref{fig_Gam_w-rho}$a$
is for a $\Delta$ at rest,
and that the Pauli blocking of the emitted nucleon
has not been taken into account
(the Pauli blocking is treated explicitly in a transport description).
As seen in fig.\ \ref{fig_Gam_w-rho}$a$
the width is quite substantial
at the quasipion energy $\hbar \tilde{\omega}_1 = 200\ \MeV$,
about $80\ \MeV$ compared to the free width,
which is about $30\ \MeV$ at this energy.
Thus it is quite probable for a $\Delta$ to decay
to the mode $\tilde{\pi}_1$ around this quasipion energy.

Apart from penetrating the surface
there is also some probability
for the mode $\tilde{\pi}_1$
to be reflected at the surface.
This is difficult to exactly predict
since the reflection coefficient, ${\cal R}$,
will depend on the actual density profile.
However a first estimate can be obtained
by considering a one-dimensional scenario
with a density profile
that corresponds to a Wood-Saxon potential
with a surface thickness $a=0.65$ fm.
The height of the potential is then given
by the change in the momentum of the pionic mode,
and for the case of $\hbar \tilde{\omega}_1 = 200\ \MeV$
we obtain  ${\cal R} \approx 0.017$,
see further the discussion in appendix \ref{sec_Refl}.
This number is very small,
which implies that the reflection
can practically be neglected for this particular case.
However, in other situations
where the momentum change is different or
the density profile is sharper,
the reflection coefficient may be larger.
We have therefore devoted appendix \ref{sec_Refl}
to discuss how effective local reflection coefficients
can be obtained and implemented in transport descriptions.

In our second example we also consider the mode $\tilde{\pi}_1$,
but now with energy $\hbar \tilde{\omega}_1 = 295\ \MeV$.
Initially this mode will now have the characteristics:
\[
\begin{array}{cccc}
q \approx 450\ \MeV/c
&
|Z|^2 \approx 0.0
&
\sum_{\Delta N^{-1}} |X_{\Delta N^{-1}}|^2 \approx 0.94
&
\sum_{N N^{-1}} |X_{N N^{-1}}|^2 \approx 0.06\ .
\end{array}
\]
As the density decreases it
follows the path indicated
by the dot-dashed line in fig.\ \ref{fig_w12-rho}$a$,
from $q \approx 450\ \MeV/c$ to $q \approx 250\ \MeV/c$.
In fig.\ \ref{fig_Amp-w}$b$ we see how the squared amplitudes
vary with the density for this particular energy
of the mode $\tilde{\pi}_1$.
As seen in fig.\ \ref{fig_Amp-w}$b$
also in this case only the pion component remains,
as the zero density limit is reached,
although $\tilde{\pi}_1$ initially was completely
dominated by the $\Delta N^{-1}$ component.
Thus also in this case
the mode will escape the system as a free pion.

In fig.\  \ref{fig_DhAmp-w}$b$
we show the squared amplitudes
of the individual $\Delta N^{-1}$ amplitudes
for the energy $\hbar \tilde{\omega}_1 = 295\ \MeV$
of the mode $\tilde{\pi}_1$.
Here the situation is different from the case at
$\hbar \tilde{\omega}_1 = 200\ \MeV$, since initially
the mode is dominated by a single $\Delta N^{-1}$ component,
\ie\ the mode is non-collective.
However as the density is lowered
the strength is spread over more $\Delta N^{-1}$ components
and at low densities the mode is completely collective.

{}From fig.\ \ref{fig_Gam_w-rho}$a$ we see that the partial width
$\tilde{\Gamma}_\Delta^1$ is very close to zero
at the quasipion energy
$\hbar \tilde{\omega}_1 = 295\ \MeV$ (dot-dashed curve)
at normal nuclear density.
Thus a $\Delta$ at normal nuclear density
will not decay to a quasipion with this energy.
The partial $\Delta$ width becomes very small
because the collective strength,
as well as the pion component,
is negligible in this case.
Although the mode at this energy
cannot be created at normal nuclear density,
it may still be created at lower densities,
as can be seen in fig.\ \ref{fig_Gam_w-rho}$a$.

Making the same assumptions as in the first example,
we find that the reflection coefficient,
becomes smaller than $10^{-4}$ for this case.

We thus conclude that the lower pionic mode
penetrating the surface will always emerge as a free pion.
For low energies this is natural
because as the density decreases
the dispersion relation of the pionic mode
approaches the free pion relation.
The cases of sufficiently large energy
to approach the unperturbed $\Delta N^{-1}$ branch
as the density is lowered
will never occur since no such modes will be created
from a decaying $\Delta$,
because the partial $\Delta$ width will be zero.

\subsection{ Upper pionic mode }
\label{sec_QpionsUpp}
Also on the upper collective mode,
$\tilde{\pi}_2$ a similar effect will occur.
However some properties are somewhat different
so we will therefore illustrate
also this case with two typical examples.
In our third example we thus consider the mode $\tilde{\pi}_2$,
with energy $\hbar \tilde{\omega}_2 = 320\ \MeV$.
Note that at this energy the mode can only exist
at densities up to about $0.5 \rho^0$.
Initially, at $\rho = 0.5 \rho^0$,
this mode will have the characteristics:
\[
\begin{array}{cccc}
q \approx 0\ \MeV/c
&
|Z|^2 \approx 0.0
&
\sum_{\Delta N^{-1}} |X_{\Delta N^{-1}}|^2 \approx 1.0
&
\sum_{N N^{-1}} |X_{N N^{-1}}|^2 \approx 0.0\ .
\end{array}
\]
As the density decreases we
follow the path indicated
by the dashed line in fig.\ \ref{fig_w12-rho}$b$,
from $q \approx 0\ \MeV/c$ to $q \approx 270\ \MeV/c$.
Note that on the upper collective mode the momentum increases as the density
decreases, corresponding to a negative potential step.

In fig.\ \ref{fig_Amp-w}$c$ we see how the squared amplitudes
vary with the density for this particular energy
of the mode $\tilde{\pi}_2$.
As seen in fig.\ \ref{fig_Amp-w}$c$,
contrary to previous examples,
only the $\Delta N^{-1}$ component remains,
as the zero density limit is reached.
In fig.\  \ref{fig_DhAmp-w}$c$
we show the squared amplitudes
of the individual $\Delta N^{-1}$ amplitudes
for the energy $\hbar \tilde{\omega}_2 = 320\ \MeV$
of the mode $\tilde{\pi}_2$.
Initially the mode is collective, but as the density is lowered the mode
becomes more and more non-collective.
{}From fig.\ \ref{fig_Gam_w-rho}$b$ we however see that the partial width
$\tilde{\Gamma}_\Delta^2$ actually is zero
at the quasipion energy
$\hbar \tilde{\omega}_2 = 320\ \MeV$ (dot-dashed curve)
at all densities.
Thus a $\Delta$
will not decay to a quasipion with this energy.

A mode at this energy with a very low momentum at half nuclear density,
which could occur in a time-dependent density,
could have a very large reflection coefficient,
approaching unity as the initial quasipion momentum approaches zero.

In our fourth and last example in this section
we consider the mode $\tilde{\pi}_2$
with energy $\hbar \tilde{\omega}_2 = 380\ \MeV$.
Initially this mode will now have the properties:
\[
\begin{array}{cccc}
q \approx 170\ \MeV/c
&
|Z|^2 \approx 0.12
&
\sum_{\Delta N^{-1}} |X_{\Delta N^{-1}}|^2 \approx 0.88
&
\sum_{N N^{-1}} |X_{N N^{-1}}|^2 \approx 0.0\ .
\end{array}
\]
As the density decreases we
follow the path indicated
by the dot-dashed line in fig.\ \ref{fig_w12-rho}$b$,
from $q \approx 170\ \MeV/c$ to $q \approx 350\ \MeV/c$.
In fig.\ \ref{fig_Amp-w}$d$ we see how the squared amplitudes
vary with the density for this particular energy
of the mode $\tilde{\pi}_2$.
As seen in fig.\ \ref{fig_Amp-w}$d$
in this case only the pion component remains,
as the zero density limit is reached,
although $\tilde{\pi}_2$ initially was
dominated by the $\Delta N^{-1}$ component.
Thus also in this case
the mode will escape the system as a free pion.

In fig.\  \ref{fig_DhAmp-w}$d$
we show the squared amplitudes
of the individual $\Delta N^{-1}$ amplitudes
for the energy $\hbar \tilde{\omega}_2 = 380\ \MeV$
of the mode $\tilde{\pi}_2$.
We see that the mode at all densities is completely collective.
{}From fig.\ \ref{fig_Gam_w-rho}$b$ we see that the partial width
$\tilde{\Gamma}_\Delta^2$ is quite substantial
at the quasipion energy
$\hbar \tilde{\omega}_2 = 380\ \MeV$ (dot-dashed curve)
also at normal nuclear density.
The reflection coefficient,
making the same assumptions as in the first example,
becomes for this case smaller than $10^{-3}$.

We thus conclude that also the upper pionic mode
penetrating the surface will always emerge as a free pion.
For high energies this is natural
because as the density decreases
the dispersion relation of the pionic mode
approaches the free pion relation.
The cases of low energy when
the unperturbed $\Delta N^{-1}$ branch is approached
as the density is lowered,
will never occur since no such modes will be created
from a decaying $\Delta$,
because the partial $\Delta$ width will be zero.

\subsection{ Refined scenarios }
\label{sec_QpionsOther}
If the surface changes with time
the energy of the pionic mode
need not to be conserved,
and there is some small possibility
for the pionic mode to end up
as an unperturbed $\Delta$-hole state.
The actual fraction of such modes is hard to estimate
without an explicit transport simulation,
but based on the scenario for the time-independent density profile,
it is reasonable to expect
that only a very small fraction of the pionic modes
will end up as unperturbed $\Delta$-hole states in vacuum.

In a quantum description
such a mode could be either reflected at the surface,
or the mode could break up into an uncoupled $\Delta$ and hole,
where the hole is trapped inside the nucleus,
and the $\Delta$ escape the system as a free $\Delta$.
Based on the results discussed
in the explicit examples of this section,
and the presentation in appendix \ref{sec_Refl}
we expect that the reflection at the surface
will be very small.

In a transport description the reflection
can be incorporated by a local transmission coefficient
as discussed in appendix \ref{sec_Refl}.
If the pionic mode is not reflected at the surface,
and its amplitude approaches 100\%
of the $\Delta N^{-1}$ component
as the density approaches zero,
the mode should thus break up
into an uncoupled $\Delta$ and hole.
In a transport simulation this could be practically handled
by allowing the pionic mode to absorb a nearby nucleon,
forming an uncoupled $\Delta$,
when the density falls below a specified value.
This prescription has some quantum mechanical justification
by the fact that at very low density
a wave packet representing the pionic mode,
will not be very well localized,
but instead have a large spatial spread.

\section{Summary}
\label{sec_Sum}
In-medium properties obtained in an infinite stationary system
consisting of interacting nucleons, nucleon resonances and mesons,
can be incorporated into transport descriptions
by a local density approximation.
While such a prescription is rather straightforward to implement
in the interior regions of the nuclear system,
conceptual problems exist at the nuclear surface.
When the nuclear density approaches zero,
collective mesonic modes formed in the medium
have to be converted to real particles in vacuum.
The problems arise since some collective modes
(e.g. $\Delta N^{-1}$-like)
may exists in the infinite
stationary system at arbitrary low (but non-vanishing) density,
but no corresponding real particle exists in vacuum.
This problem has been apparent in previous works \cite{Giessen,Texas}
where collective modes have been incorporated
into transport descriptions as quasimesons.
The character of the quasimesons (\ie\ realization in vacuum)
were in those works determined already at the time of creation.

Based on the formalism of ref.\ \cite{main},
we have in this paper employed a more elaborate
$\pi + N N^{-1} + \Delta N^{-1}$ model
(relative to the works \cite{Giessen,Texas})
to investigate a somewhat different treatment
of the collective pionic modes at a nuclear surface.
In this formalism we have obtained
not only density dependent dispersion relations
of the pionic modes,
but also density dependent amplitudes
of the components constituting the pionic mode.
These quantities are conveniently parametrized
with density dependent parameters,
in section \ref{sec_DispAmp}.

For the transport process it is not needed
to determine the character of the pionic modes
until they penetrate the surface and emerge as free particles.
This is automatically determined
within our formalism from the amplitudes at zero density.
We have further showed in section \ref{sec_Qpions}
that for a stationary density profile,
the conservation of the energy of the pionic mode
and the partial $\Delta$ decay width,
together leads to the fact
that only real pions are realized as free particles
when the pionic mode penetrates the surface.
Note that this finding is different from earlier works,
and it demonstrates the importance
of deriving dispersion relations and partial $\Delta$ widths
consistently within
a realistic\footnote{By ``realistic'' we here mean
                     that the $\Delta N^{-1}$ model
                     contains a continuum of
                     $\Delta N^{-1}$ and $N N^{-1}$ states,
                     as compared to the more simple
                     two-level $\Delta N^{-1}$ model
                     used for example in refs.\ \cite{Giessen,Texas}.
                    }
$\Delta N^{-1}$ model.

In a more refined scenario,
where the density changes with time,
deviations from this picture can be expected
and also the unperturbed $\Delta N^{-1}$ component
may be realized in the limit of vanishing density.
The actual fraction of such modes
is hard to estimate without an explicit transport simulation,
but based on the arguments in section \ref{sec_Qpions}
for the stationary surface,
we expect this fraction to be very small.

For the rare cases when the unperturbed $\Delta N^{-1}$ component
is realized in the limit of vanishing density,
the pionic mode must be converted to real particles.
In an extended description this could be made
by allowing the mode to break up into
an uncoupled $\Delta$ and a $N^{-1}$,
where the hole remains trapped in the nuclear system
and the $\Delta$ escapes the system.
Based on our formalism,
this seems to be the most probable scenario,
since the collective strength disappears on these modes
as the density approaches zero.
This could be implemented in a transport simulation
by letting the pionic mode absorb a nearby nucleon
to form an uncoupled $\Delta$,
as may be justified by the quantum-mechanical feature that
the wave packet representing the pionic mode is not well localized
at very low density.

Alternatively the pionic modes could in a quantal description
be reflected at the surface.
We have investigated
the reflection and transmission probabilities
for the collective modes
in a simplified one-dimensional scenario,
where the modes propagate perpendicular to the surface.
Exploring typical scenarios, we have found
that (with only very few exceptions)
the reflection of the pionic modes will be smaller than a few percent,
however, in appendix \ref{sec_Refl} we have
suggested how the reflection and transmission probabilities
could be incorporated into the transport descriptions
by using approximative local transmission coefficients. \\

\noindent
Stimulating discussions with Volker Koch are acknowledged.
This work was supported by the Swedish Natural Science Research Council,
and by the Director, Office of Energy Research, Office of High Energy
and Nuclear Physics, Nuclear Physics Division of the U.S. Department of
Energy under Contract No.\ DE-AC03-76SF00098.

\appendix
\section{Reflection and transmission at a surface}
\label{sec_Refl}
In the semi-classical transport descriptions
all (test)particles are treated as classical particles.
When such a classical particle propagates
in a spatially varying potential its velocity will change.
But assuming its energy exceeds
the maximum value of the potential,
the particle will continue to propagate
through the varying potential.
This is in contrast to a quantal description
where the particle is represented by a wave packet.
This wave packet has some probability
to be reflected in a spatially varying potential.
The reflection probability depends on several factors,
such as the energy of the wave packet,
and the height and shape of the potential.

In this section we investigate
reflection and transmission probabilities
in different idealized situations
for a one-dimensional scenario,
corresponding to the direction normal to the nuclear surface.
We will also discuss how these effects
could be approximately incorporated into transport descriptions.
The treatment is intended to be used
for the collective spin-isospin modes (quasi-pions)
approaching a nuclear surface.
The formalism, though, is quite general
and could be applied also for other particles.

We will start to investigate some special cases
when the potential is stationary.
Subsequently we will discuss
how the non-stationary case could be treated.
In all cases we will assume that the potential
is constant outside a finite interval,
\begin{equation}
  V(x) = \left\{
         \begin{array}{ll}
           0    & x < x_L \\
           V(x) & x_L < x < x_R \\
           V_0  & x_R < x
         \end{array}
         \right. \ .
\end{equation}
For $x < x_L$ we have an incoming wave
and we ask for the probability
that we have an outgoing wave at $x > x_R$,
\ie\ we seek the transmission and
reflection coefficients,
${\cal T}$ and ${\cal R}$ respectively,
with ${\cal T}+{\cal R}=1$.
Only for some special simple forms of the potential $V(x)$
can the coefficients ${\cal R}$ and ${\cal T}$ be obtained analytically.

Note that both the stationary Schr\"{o}dinger equation
\begin{equation}
    \left[ -\frac{\hbar^2}{2m} \frac{d^2}{dx^2} + V(x) \right] \psi(x)
=
    E \psi(x)
\label{eq_sta-SE}
\end{equation}
and the stationary Klein-Gordon equation
\begin{equation}
    \left[ \hbar^2 c^2 \frac{d^2}{dx^2} + \hbar^2 \omega^2 -
           m^2 c^4 \right] \psi(x)
=
    2 \hbar \omega V(x) \psi(x)
\label{eq_sta-KG}
\end{equation}
for a particle with energy $E=\hbar \omega$,
can be rewritten in the form
\begin{equation}
    \left[ \frac{d^2}{dx^2} + k(x)^2 \right] \psi(x) = 0\ .
\label{eq_sta-SE+KG}
\end{equation}
by introducing a local wave number
\begin{equation}
   k(x) = \sqrt{2 m [E - V(x)]}/\hbar \ ,
\label{eq_k(x)-SE}
\end{equation}
or
\begin{equation}
     k(x) = \sqrt{\hbar^2 \omega^2 - m^2 c^4 -
                  2 \hbar \omega V(x)]}/\hbar c\ ,
\label{eq_k(x)-KG}
\end{equation}
respectively.
The results obtained in the sequel in this section
are derived using the Schr\"{o}dinger equation,
but from eqs.\ (\ref{eq_sta-SE}) to (\ref{eq_sta-SE+KG})
it follows that the results are valid also
for a particle described by the Klein-Gordon equation,
if the local wave number is taken according
to equation (\ref{eq_k(x)-KG}) instead
of equation (\ref{eq_k(x)-SE}).

The simplest case is a potential step at $x=0$,
\begin{equation}
  V(x) = \left\{
         \begin{array}{ll}
           0    & x < 0 \\
           V_0  & 0 < x
         \end{array}
         \right. \ .
\label{eq_PotStep}
\end{equation}
This case can be found in elementary textbooks on quantum mechanics.
Here we briefly recapitulate the main steps
as a preparation for more complicated cases.
The Schr\"{o}dinger equation has the solution
\begin{equation}
  \psi(x) = \left\{
         \begin{array}{ll}
           \frac{A}{\sqrt{k_L}} {\rm e}^{i k_L x} +
           \frac{B}{\sqrt{k_L}} {\rm e}^{- i k_L x}     & x < 0 \\
           \frac{C}{\sqrt{k_L}} {\rm e}^{i k_R x} +
           \frac{D}{\sqrt{k_L}} {\rm e}^{- i k_R x}     & 0 < x
         \end{array}
         \right. \ ,
\end{equation}
where $k_L = \sqrt{2 m E}/\hbar $
and $k_R = \sqrt{2 m (E-V_0)}/\hbar $,
for a particle with energy $E>V_0$.
Taking only an outgoing solution at $x > x_R$ (\ie\ $D=0$),
and relating the coefficients $A$, $B$ and $C$
by a smooth joining at $x=0$,
\begin{equation}
  \left\{
         \begin{array}{l}
           \psi(0-\epsilon) = \psi(0+\epsilon)    \\
           \psi'(0-\epsilon) = \psi'(0+\epsilon)    \\
         \end{array}
         \right. \ ,
\end{equation}
the reflection and transmission coefficients are obtained as
\begin{eqnarray}
   {\cal R} & = & \frac{|B|^2}{|A|^2} = \frac{(k_L-k_R)^2}{(k_L+k_R)^2}
\nonumber \\
   {\cal T} & = & \frac{|C|^2}{|A|^2} =
                  \frac{4 k_L k_R}{(k_L+k_R)^2} \ .
\end{eqnarray}

Also for other special forms of the potential
can the reflection and transmission coefficients
be derived analytically,
such as for the Woods-Saxon type of potential,
\begin{equation}
    V(x) = V_0/[1+{\rm e}^{-x/a}]\ .
\label{eq_WS-pot}
\end{equation}
Here the reflection coefficient ${\cal R}$
is given by \cite{Lan-Lif}
\begin{equation}
   {\cal R} = \left( \frac{\sinh[a \pi (k_L-k_R)]}
                   {\sinh[a \pi (k_L+k_R)]}
       \right)^2\ ,
\label{eq_WS-R}
\end{equation}
and the transmission coefficient
is obtained from ${\cal T} = 1 - {\cal R}$.
Note that the case of the potential step,
eq.\ (\ref{eq_PotStep}),
emerges in the limit $a \rightarrow 0$.

For an arbitrary potential
the coefficients ${\cal R}$ and ${\cal T}$
do not have an analytical form.
An approximate solution can be obtained
by approximating the potential $V(x)$
by a piecewise constant potential
\begin{equation}
  V(x) = \left\{
         \begin{array}{ll}
           0    & x < x_0 = x_L                              \\
           V(x) \approx \sum_{j=0}^{n-1} V_j(x)
                & x_0 = x_L < x < x_n = x_R                  \\
           V_0  & x_N = x_L < x
         \end{array}
         \right. \ ,
\end{equation}
with
\begin{equation}
        V_j(x)
=       V(x_j + \Delta x/2) [\theta(x_{j+1}-x) - \theta(x_j-x)],
\quad
        \left\{ \begin{array}{ccc}
                    x_j     & = &   x_L + j \: \Delta x \\
                  \Delta x  & = &       [x_R-x_L]/n
        \end{array} \right. \ ,
\end{equation}
and by making the ansatz,
\begin{equation}
    \psi(x) \approx \sum_{j=0}^{n-1} \psi_j(x)
                        [\theta(x_{j+1}-x) - \theta(x_j-x)]
    \qquad  x_0 = x_L < x < x_n = x_R\ ,
\end{equation}
with
\begin{equation}
    \psi_j(x)
=
    \frac{A_j}{\sqrt{k_j}} {\rm e}^{i k_j x} +
    \frac{B_j}{\sqrt{k_j}} {\rm e}^{-i k_j x}
\end{equation}
and
\begin{equation}
   k_j = \sqrt{2 m [E - V(x_j + \Delta x/2)]}\hbar \ .
\end{equation}
The coefficients $A_j$ and $B_j$ are determined
from the condition of smooth joining,
\begin{equation}
  \left\{
         \begin{array}{l}
           \psi_j(x_{j+1}-\epsilon)  = \psi_{j+1}(x_{j+1}+\epsilon)     \\
           \psi_j'(x_{j+1}-\epsilon) = \psi_{j+1}'(x_{j+1}+\epsilon)    \\
         \end{array}
         \right. \ ,
         \qquad \epsilon \rightarrow 0^+\ ,
\end{equation}
and the boundary condition
of an outgoing solution at $x>x_R$,
\ie\ $A_{n-1} = 1$ and $B_{n-1}=0$.
Not that in the especially simple case of no reflection,
the ansatz above becomes identical to the WKB approximation,
in which the the coefficients $A_j$ and $B_j$ are constants,
\ie\  $A_j \equiv A$ and $B_j \equiv B$ for all $j$.

The transmission coefficient, ${\cal T}$, is obtained from
\begin{equation}
   {\cal T}  = \frac{|A_{n-1}|^2}{|A_0|^2}\ ,
\end{equation}
and ${\cal R} = 1 - {\cal T}$.
We have numerically checked
that for the Woods-Saxon type of potential above,
this approximation procedure,
with a very high accuracy,
yields the same reflection and transmission coefficients
as the analytical result.
We have compared the analytical and numerical results
for many different energies ($E$)
and parameters of the potential ($V_0$, $a$),
and only for $E$ very close to $V_0$
are there deviations larger than a few percent.

In a heavy-ion collision the potential is not stationary,
but changes with the time.
The formalism above can be generalized
to a time-dependent potential.
The particles to be propagated
are in a quantal description
represented by wave packets.
We therefore generalize the definition
of the transmission coefficient to be valid
also for a propagating wave packet in a non-stationary potential.
As before we assume that we have
an incident wave or wave packet,
$\psi_{\rm inc}(x,t)$, at $x=x_L$.
We associate a probability current with this wave packet by
\begin{equation}
    j_{\rm inc}(x,t)
=
    \frac{\hbar}{m} \mbox{ Im }
    \left( \frac{d \psi_{\rm inc}(x)}{dx} \; \psi_{\rm inc}(x)^* \right)
\end{equation}
Similarly we assume that we have
an outgoing wave or wave packet,
$\psi_{\rm out}(x,t)$,  at
$x = x_R$,
and we associate a probability current,
$j_{\rm out}(x,t)$,
with this wave packet.
A transmission coefficient can then be defined
as the ratio between the outgoing probability
flowing through the point $x_R$
and the incident probability flowing through the point $x_L$,
\begin{equation}
    {\cal T} = \frac{ \int_{t_1}^{t_2} j_{\rm out}(x_R,t) dt }
                    { \int_{t_1}^{t_2} j_{\rm inc}(x_L,t) dt }\,
\label{eq_TrCo-t}
\end{equation}
where the times $t_1$ and $t_2$ are chosen such
that a wave packet will propagate through
both the points $x_L$ and $x_R$ within the time interval.
Note that this definition of ${\cal T}$
agrees with the results stated for the stationary case.
By solving the time-dependent Schr\"{o}dinger equation numerically,
and expanding a wave packet in this basis set,
a transmission coefficient could in principle
be obtained from eq. (\ref{eq_TrCo-t})
for an arbitrary (known) potential $V(x,t)$.
In a transport description, however,
the potential $V(x,t)$ is not known in advance,
and the described method is not so useful.
Instead it would be more useful
with a local transmission coefficient $t(x_j)$,
which would give the probability for the wave packet
to be transmitted from the point $x_j$
to the point $x_{j+1} = x_j + \Delta x$,
where the total transmission coefficient is given by
${\cal T} = \prod_{j=0}^{n-1} t(x_j)$.
Such a local coefficient can formally be obtained by writing
\begin{equation}
    {\cal T} = {\rm e}^{ - \int_{x_L}^{x_R} r(x) dx}\ ,
\label{eq_T(x)}
\end{equation}
and by approximating the integral in the exponent by a discrete sum,
\begin{equation}
         {\cal T}
\approx
         {\rm e}^{ - \sum_{j=0}^{n-1} r(x_j) \Delta x}
=
         \prod_{j=0}^{n-1} {\rm e}^{ - r(x_j) \Delta x}
=
         \prod_{j=0}^{n-1} t(x_j)\ .
\label{eq_Ttot-loc}
\end{equation}
However, also in this expression
the local reflection coefficient $r(x)$
will depend on the particle energy,
and the shape and magnitude of the potential,
and it is therefore not straightforward
to implement a local transmission coefficient
into a transport description.

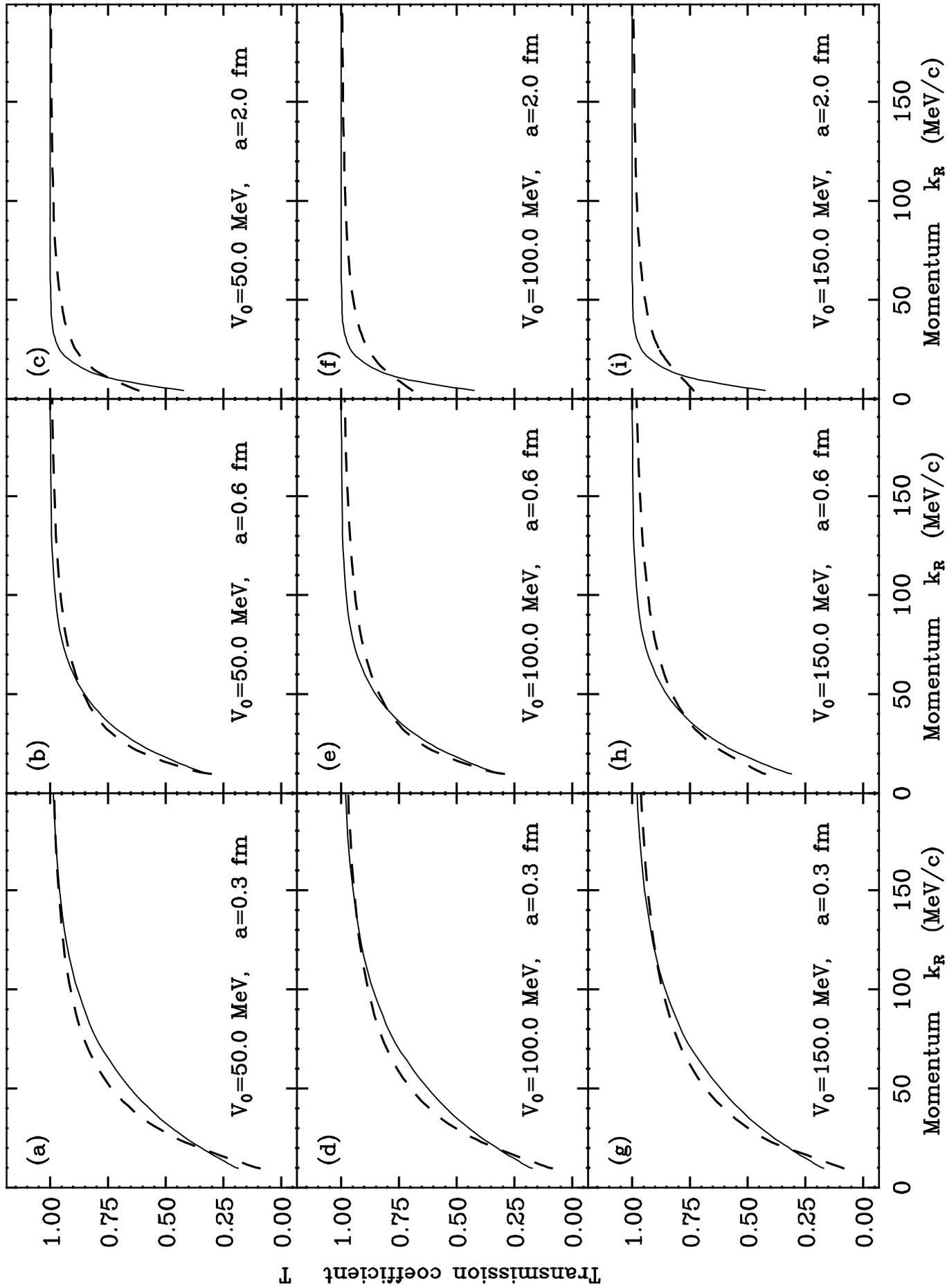
\begin{figure}\caption{}\label{fig_LocT}
Total transmission coefficient, ${\cal T}$,
obtained for a Woods-Saxon type of potential,
eq.\ (\ref{eq_WS-pot}),
for a range of potential parameters $V_0$ and $a$.
The solid curve represents
the exact expression, eq.\ (\ref{eq_WS-R}),
while the dashed curve represents the approximative result
obtained from the local transmission coefficient,
eqs.\ (\ref{eq_Ttot-loc}) and (\ref{eq_tloc}).
\end{figure}

It is therefore our strategy
to approximate the function $r(x)$
with a relative simple expression
depending on a few parameters,
that are adjusted
such that eq. (\ref{eq_T(x)})
will yield a good approximation
to the total transmission coefficient
for a large class of realistic particle energies
and potential parameters.
We have empirically found that the ansatz
\begin{equation}
    r(x)
=
    \frac{C_1}{k(x)}
    \left( \frac{d \ln k(x) }{dx} \right)^2\ .
\end{equation}
with $C_1 \approx 0.116$,
yields a good approximation for a range
of different particle energies
and a range  of different values of $a$ and $V_0$
in the Woods-Saxon type of stationary potential.
In fig.\  \ref{fig_LocT} we present
some examples of this approximation,
compared to the analytical solutions.

As was discussed in section \ref{sec_Qpions}
it is only in rather few and special cases
that the transmission coefficient
for the collective pionic modes
will deviate substantially from unity.
We therefore conclude this section
by summarizing that the reflection of pionic modes
can be approximately incorporated into transport descriptions
by an approximate local transmission coefficient
\begin{equation}
         t(x)
\approx
         \exp[ - \frac{C_1}{k(x)}
                 \left( \frac{d \ln k(x) }{dx} \right)^2
                 \Delta x]\ .
\label{eq_tloc}
\end{equation}
The error in this approximation
is in most realistic cases smaller than 10\%.
Those events where a larger error may occur
for pionic modes at a nuclear surface
are expected to be very rare in transport simulations.
Thus the total error in a transport treatment
by using a local transmission coefficient
according to eqs.\ (\ref{eq_Ttot-loc}) and (\ref{eq_tloc})
should be very small.
However, the importance of incorporating the effects
of reflection of pionic modes  at the surface
seems to be quite small,
since in most events the total transmission coefficient
will be close to unity.

\section{ RPA equations for spin-isospin interaction }
\label{sec_RPAsolu}

In this section we recapitulate the RPA equations,
derived in ref.\ \cite{main},
in the spin longitudinal channel
for the case when $T=0$
and the $\Delta$ width is omitted in the RPA equations.

The spin-isospin excitations are characterized
by the momentum $\mathbf{q}$ and the isospin $\lambda$.
The energy of the spin-isospin modes
are obtained from the determinant of the equation
\begin{equation}
  \left(
         \begin{array}{cc}
              1 + {\cal W}^{NN} {\cal M}^N \Phi^N
         &    {\cal W}^{N \Delta} {\cal M}^\Delta \Phi^\Delta       \\
              {\cal W}^{\Delta N} {\cal M}^N \Phi^N
         &    1 + {\cal W}^{\Delta \Delta} {\cal M}^\Delta \Phi^\Delta
         \end{array}
  \right)
  \left(
         \begin{array}{c}
           x^N \\ x^\Delta
         \end{array}
  \right) = 0\ ,
\label{eq_Xx}
\end{equation}
while the amplitudes are obtained
from the solution of eq.\ (\ref{eq_Xx}),
\begin{eqnarray}
  Z & = &
      \frac{1}{\hbar \omega_\pi - \hbar \omega}
      \left[ {\cal M}^N v_\pi^N \Phi^N x^N +
             {\cal M}^\Delta v_\pi^\Delta \Phi^\Delta x^\Delta \right]
      \label{eq_Zx}  \\
  W & = &
      \frac{1}{\hbar \omega_\pi + \hbar \omega}
      \left[ {\cal M}^N v_\pi^N \Phi^N x^N +
             {\cal M}^\Delta v_\pi^\Delta \Phi^\Delta x^\Delta \right]\ ,
      \label{eq_Wx}
\end{eqnarray}
and
\begin{equation}
  X_{jk}(\omega,\mathbf{q},\lambda) =
     \left( \frac{\hbar c}{L} \right)^{3/2}
     \frac{x(t_j,t_k; \, \mathbf{q}; \, \lambda, \, \omega)}
          {e_{t_j}(\mathbf{p}_k + \mathbf{q}) - e_{t_k}(\mathbf{p}_k)
           + \Sigma_{jk} -
           \hbar \omega} \; \vartheta_{j k}(\hat{q},-\lambda)\ ,
\label{eq_Xansatz}
\end{equation}
with
\begin{eqnarray*}
 x(1/2,1/2; \, \mathbf{q}; \, \lambda, \, \omega)
&
  \equiv
&
         x^N
\\
 x(1/2,3/2; \, \mathbf{q}; \, \lambda, \, \omega)
&
  =
&
         x(3/2,1/2; \, \mathbf{q}; \, \lambda, \, \omega)
  \equiv x^{\Delta}
\\
 x(3/2,3/2; \, \mathbf{q}; \, \lambda, \, \omega)
&
  \equiv
&
         0
\end{eqnarray*}
and analogously for other quantities.
In the above expressions $\alpha,\beta = N,\Delta$,
and we have used the notations
\begin{eqnarray}
  {\cal W}^{\alpha \beta} & = &
       \frac{f^\pi_{N \alpha} f^\pi_{N \beta}}{(m_\pi c^2)^2}
       [ \: |F_g|^2 g_{\alpha \beta}' +
         R^\alpha_i R^\beta_i
         |F_\pi|^2
         (c \mathbf{q}_i)^2 D^0_\pi\: ], \\
  v_{\pi}^\alpha(\mathbf{q},\omega)
 & = &
       i \, R^\alpha_i
       F_{\pi}
       \frac{ f^\pi_{N \alpha} }{m_{\pi} c^2}
       \frac{|c \mathbf{q}_{cm}|}
            {\sqrt{2 \hbar \omega_{\pi}(\mathbf{q})}}      \\
  R_\alpha^i(q)^2
& = & \frac{ 2m_\alpha c^2}{m_\alpha c^2 + \sqrt{s_i} } \\
  \sqrt{s}
 & =  &
        (E_N(\mathbf{p}_N) + \hbar \omega)^2 -
             (c \mathbf{q} + c \mathbf{p}_N)^2  \\
 & \approx &
        (m_N c^2 + \mbox{ Re } \hbar \omega)^2 - c(\mathbf{q})^2
         \equiv \sqrt{s_i} \\
  \mathbf{q}_{cm}
 & \approx &
  \mathbf{q}_i = \frac{m_N c^2} {m_N c^2 + \hbar \omega} \mathbf{q}\ .
\label{eq_vpi}
\end{eqnarray}
The numerical factors
${\cal M}^N = 4$ and ${\cal M}^\Delta = 16/9$
originates from the spin-isospin summation,
and we have defined the Lindhard functions
\begin{eqnarray}
  \Phi^N(\omega, \mathbf{q}) & = &
     \left( \frac{\hbar c}{L} \right)^3 \sum_{\mathbf{p}}
     \frac{n(\mathbf{p}) - n(\mathbf{p} + \mathbf{q})}
          { (\mathbf{p} + \mathbf{q})^2/ 2 m_N^*  -
             \mathbf{p}^2/2 m_N^* - \hbar \omega}
\label{eq_PhiN} \\
  \Phi^\Delta(\omega, \mathbf{q}) & = &
     \Phi(\frac{1}{2},\frac{3}{2}; \: \omega, \mathbf{q}) +
     \Phi(\frac{3}{2},\frac{1}{2}; \: \omega, \mathbf{q}) =
     \left( \frac{\hbar c}{L} \right)^3 \sum_{\mathbf{p}}
     \left\{ \frac{n(\mathbf{p})}{\delta e^+_{\Delta N} } +
             \frac{n(\mathbf{p})}{\delta e^-_{\Delta N} } \right\}
\label{eq_PhiD}
\end{eqnarray}
with
\begin{eqnarray}
  \delta e^+_{\Delta N} & = &
          \frac{ (\mathbf{p}_\Delta^+)^2 }{ 2 m_\Delta } -
            \frac{ \mathbf{p}^2}{ 2 m_N^* } + \Delta m
            + V_\Delta^{\rm eff}(\varepsilon_\Sigma^+, \mathbf{p}_\Delta^+)
            - \hbar \omega  \\
  \delta e^-_{\Delta N} & = &
           \frac{ (\mathbf{p}_\Delta^-)^2 }{ 2 m_\Delta } -
            \frac{ \mathbf{p}^2}{ 2 m_N^* } + \Delta m
            + V_\Delta^{\rm eff}( \varepsilon_\Sigma^-, \mathbf{p}_\Delta^-)
            + \hbar \omega
\end{eqnarray}
\begin{equation}
  \Delta m = m_\Delta - m_N; \qquad
  \varepsilon_\Sigma^\pm =
        m_N + \frac{\mathbf{p}^2}{2 m_N^*} \pm \hbar \omega; \qquad
  \mathbf{p}_\Delta^\pm = \mathbf{p} \pm\ \mathbf{q}\ .
\end{equation}

The quantities $x^N$ and $x^{\Delta}$ are determined
from the normalization of the RPA states
\begin{eqnarray}
       \pm 1
& = &
       {\cal M}^N \eta_N  x_N^* x_N
+
       {\cal M}^\Delta \eta_\Delta  x_\Delta^* x_\Delta
+
       Z^* Z - W^* W
\nonumber \\
& = &
       x_N^* x_N
       \left( {\cal M}^N \eta_N
+
              \left[ \frac{ 1 + {\cal W}^{NN} {\cal M}^N \Phi^N }
                          { {\cal W}^{N \Delta} {\cal M}^\Delta \Phi^\Delta }
              \right]^2 {\cal M}^\Delta \eta_\Delta
       \right.
\\
& - &
      \left.
              \left[ \frac{1}{(\hbar \omega_\pi - \hbar \omega)^2} -
                     \frac{1}{(\hbar \omega_\pi + \hbar \omega)^2}
              \right]
              \left\{ {\cal M}^N v_\pi^N \Phi^N
-
                      \frac{ 1 + {\cal W}^{NN} {\cal M}^N \Phi^N }
                           { {\cal W}^{N \Delta} } v_\pi^\Delta
              \right\}^2
      \right)
\nonumber
\label{eq_SoNormzFi}
\end{eqnarray}
with
\begin{eqnarray}
 \eta_N(\omega, \mathbf{q})
&
  =
&
   \left( \frac{\hbar c}{L} \right)^3 \sum_{\mathbf{p}}
   \frac{n(\mathbf{p}) - n(\mathbf{p} + \mathbf{q})}
        { [ \frac{ (\mathbf{p} + \mathbf{q})^2 }{ 2 m_N^* } -
            \frac{ \mathbf{p}^2}{ 2 m_N^* } - \hbar \omega]^2 }
 =
   \frac{\partial}{\partial \hbar \omega} \Phi_N(\omega, \mathbf{q})
\\
 \eta_\Delta(\omega, \mathbf{q})
&
 =
&
   \left( \frac{\hbar c}{L} \right)^3 \sum_{\mathbf{p}} \left\{
   \frac{n(\mathbf{p})}{ [ \delta e^+_{\Delta N} ]^2} -
   \frac{n(\mathbf{p})}{ [ \delta e^-_{\Delta N} ]^2} \right\}
 =
   \frac{\partial}{\partial \hbar \omega} \Phi_\Delta(\omega, \mathbf{q})\ ,
\end{eqnarray}
and where we have assumed
$\partial V_\Delta^{\rm eff}/\partial \hbar \omega \approx 0$.

The factor $h(jk;\nu)$,
motivated in eq.\ (\ref{eq_hMotiv}),
can using the RPA solution
be explicitly expressed as
\begin{eqnarray}
 h(jk,\nu;\omega,\mathbf{q})  & = &
   \sum_{\alpha=N,\Delta}
   {\cal M}_\alpha \Phi_\alpha(\omega_\nu) x_\alpha(\omega_\nu)
    \left[ (\hat{\mathbf{q}}_{jk} \mathbf{\cdot} \hat{\mathbf{q}}_i)
            v_B^{\alpha,jk}(\omega)
    \right. \nonumber \\ & & \left.
            -\ 2 \hbar \omega_{\pi}
            D_{\pi}^0(\omega_\nu)
            v_{\pi}^{jk}(\omega) v_{\pi}^{\alpha}(\omega_\nu)
            \right]\ ,
\end{eqnarray}
where
\begin{equation}
  v_B^{\alpha,jk}(\omega,\mathbf{q}) = g'_{\alpha,jk}
      \frac{f^\pi_{N,\alpha} f^\pi_{N,jk} }{(m_\pi c^2)^2}\
      F_g^2(\omega,\mathbf{q}) \ .
\end{equation}

\newpage
\begin{table}[t]
\begin{tabular}{||l|l|l||}
  \hline
        $m_N = 940\ \MeV/c^2$
&       $g'_{NN} = 0.9$
&       $f^{\pi}_{NN} = 1.0$  \\
        $m_{\Delta} = 1230\ \MeV/c^2$
&       $g'_{N \Delta} = 0.38$
&       $f^{\pi}_{N \Delta } = 2.2$    \\
        $m_{\pi} = 140\ \MeV/c^2$
&       $g'_{\Delta \Delta} = 0.35$
&       $f^{\pi}_{\Delta \Delta } = 0$ \\
        $\Lambda_g = 1.5\ \GeV$
&       $\Lambda^{\pi} = 1.0\ \GeV$
&        $\rho_0 = 0.153\ \fm^{-3}$  \\
  \hline
        $V_\Delta - V_N = 25.0\rho/\rho_0\ \MeV$
&       \multicolumn{2}{|l||}{$m_N^*= m_N/[1+0.4049(\rho/\rho_0)]$}  \\
  \hline
\end{tabular}
\caption{ Parameter values used in the numerical calculations.
          \protect\label{tab_param}
}
\end{table}


\newpage
\begin{figure}\caption{}\label{fig_DseGraph}
Diagrammatic representations
of the $\Delta$ self energy $\Sigma_\Delta$ (left-hand side)
and the partial $\Delta$ decay width
to the particular spin-isospin mode $\nu$,
$\Gamma_\Delta^\nu$ (right-hand side).
\end{figure}

\begin{figure}\caption{}\label{fig_DispT0r10}
The dispersion relations for the spin-isospin modes
in the spin-longitudinal channel,
in infinite nuclear matter
at normal nuclear density and zero temperature.
The non-collective modes are shown by solid curves,
while collective modes $\tilde{\pi}_1$ and $\tilde{\pi}_2$
are represented by dot-dashed curves.
As a reference, the free pion dispersion relation
$\hbar \omega_\pi(q) = [(m_\pi c^2)^2 + (cq)^2]^{1/2}$,
and the unperturbed $\Delta N^{-1}$ relation
$\hbar \omega_{\Delta N^{-1}}(q) = m_\Delta c^2 - m_N c^2 + q^2/2m_\Delta$,
are included as a dotted curves.
\end{figure}

\begin{figure}\caption{}\label{fig_DispParam}
The density-dependent parameters
$U_0$ (solid curve), $q_0$ (dashed curve) and $m_0$ (dot-dashed)
used in eq.\ (\ref{eq_DispParam})
for fitting the dispersion relations
of the lower collective mode $\tilde{\pi}_1$ (part $a$)
and the upper collective mode $\tilde{\pi}_2$ (part $b$).
\end{figure}

\begin{figure}\caption{}\label{fig_GamParNPB}
The total $\Delta$ width $\Gamma_\Delta^{\rm tot}$
and its partial contributions from different spin-isospin modes
for a $\Delta$ with momentum 300 MeV/$c$.
The solid curve represents the total width,
the long-dashed line is the contribution
from the non-collective $N N^{-1}$ modes,
the short-dashed line is the contribution
from the non-collective $\Delta N^{-1}$ modes,
the dot-dashed line is the contribution
from the lower pionic mode $\tilde{\pi}_1$,
and  the dot-dot-dashed line is the contribution from
the upper pionic mode $\tilde{\pi}_2$.
The intermediate nucleon in the $\Delta$ decay is not Pauli blocked.
The error bars indicate the estimated uncertainty associated with
the classification procedure (see ref.\ \cite{main}).
\end{figure}

\begin{figure}\caption{}\label{fig_AmplParam}
The density-dependent parameters,
$C_1$ (dot-dashed curve),
$C_2$ (short-dashed),
$C_3$ (dot-dot-dashed) and
$C_4$ (long-dashed),
for the amplitudes
of the $\Delta N^{-1}$ (part $a$) and the $N N^{-1}$ ($b$) components
on the lower pionic branch,
eq.\ (\ref{eq_XDh-low-parm}) and (\ref{eq_XNh-low-parm}).
\end{figure}

\end{document}